\documentclass{aastex63}
\usepackage{amsmath}
\usepackage{xcolor}
\usepackage{soul}
\usepackage{subfigure}
\usepackage{todonotes}
\usepackage{mhchem}
\usepackage{isotope}

\newcommand{\rp}{\emph{r}-process}
\newcommand{\Cf}{{$^{254}$Cf}}

\shorttitle{KN Light Curves: Nuclear Inputs}
\begin{document}
\title{Modeling Kilonova Light Curves: Dependence on Nuclear Inputs}

\author[0000-0003-0245-827X]{Y.~L. Zhu}
\affiliation{Department of Physics, North Carolina State University, Raleigh, NC 27695 USA}
\affiliation{Joint Institute for Nuclear Astrophysics - Center for the Evolution of the Elements, USA}

\author[0000-0003-0031-1397]{K. Lund}
\affiliation{Department of Physics, North Carolina State University, Raleigh, NC 27695 USA}

\author[0000-0003-3340-4784]{J. Barnes}
\affiliation{Department of Physics and Columbia Astrophysics Laboratory, Columbia University, New York, NY 10027, USA.}
\affiliation{Einstein Fellow}

\author[0000-0002-4375-4369]{T.~M. Sprouse}
\affiliation{University of Notre Dame, Notre Dame, Indiana 46556, USA}

\author[0000-0002-3305-4326]{N. Vassh}
\affiliation{University of Notre Dame, Notre Dame, Indiana 46556, USA}

\author[0000-0001-6811-6657]{G.~C. McLaughlin}
\affiliation{Department of Physics, North Carolina State University, Raleigh, NC 27695 USA}
\affiliation{Joint Institute for Nuclear Astrophysics - Center for the Evolution of the Elements, USA}

\author[0000-0002-9950-9688]{M.~R. Mumpower}
\affiliation{Theoretical Division, Los Alamos National Laboratory, Los Alamos, NM, 87545, USA}
\affiliation{Center for Theoretical Astrophysics, Los Alamos National Laboratory, Los Alamos, NM, 87545, USA}
\affiliation{Joint Institute for Nuclear Astrophysics - Center for the Evolution of the Elements, USA}

\author[0000-0002-4729-8823]{R. Surman}
\affiliation{University of Notre Dame, Notre Dame, Indiana 46556, USA}
\affiliation{Joint Institute for Nuclear Astrophysics - Center for the Evolution of the Elements, USA}

\begin{abstract}
The mergers of binary neutron stars, as well as black hole-neutron star systems, are expected to produce an electromagnetic counterpart that can be analyzed to infer the element synthesis that occurred in these events. 
We investigate one source of uncertainties pertinent to lanthanide-rich outflows: the nuclear inputs to rapid neutron capture nucleosynthesis calculations. 
We begin by examining thirty-two different combinations of nuclear inputs: eight mass models, two types of spontaneous fission rates and two types of fission daughter product distributions.
We find that such nuclear physics uncertainties typically generate at least one order of magnitude uncertainty in key quantities such as the nuclear heating (one and a half orders of magnitude at one day post-merger), the bolometric luminosity (one order of magnitude at five days post-merger), and the inferred mass of  material from the bolometric luminosity (factor of eight when considering the eight to ten day region).
Since particular nuclear processes are critical for determining the electromagnetic signal, we provide tables of key nuclei undergoing $\beta$-decay, $\alpha$-decay, and spontaneous fission important for heating at different times, identifying decays that are common among the many nuclear input combinations.
\end{abstract}


\section{Introduction}\label{introduction}
The recent detection of both gravitational waves and electromagnetic emission from the neutron star merger (NSM) GW170817~\citep{Abbott2017GW1708, Abbott2017multi, Dnaz2017Observ} provides us with an unprecedented wealth of data~\citep{Chornock2017TheEl,Cowperthwaite2017TheEl,NichollTheEl}, which can be used to enhance our understanding of many physical processes that accompany compact object mergers.
One such physical process is the nucleosynthesis triggered in merger-driven outflows, which has long been suspected to include heavy elements formed via rapid neutron capture (\rp)~\citep{Lattimer1974Black-,Eichler1989Nucleo, Freiburghaus1999r-Proc, Pian2017Spectr}.
As first pointed out by \citet{li1998trans}, the radioactive decay of these freshly synthesized unstable \rp{} nuclei should power an electromagnetic transient, observations of which could offer unique insight into mass ejection from mergers and astrophysical nucleosynthesis. 
In the years leading up to the detection of GW170817, significant theoretical progress was made towards understanding these macronova~\citep{Kulkarni2005Modeli}, or kilonova (KN)~\citep{Metzger2010Electr} signals (for recent reviews, see \citet{Nakar2019Theel, Metzger2020Kilono}.)

The process of modeling and interpreting KN emissions encompasses multi-physics problems that require tracking astrophysical, nuclear and thermodynamic processes.
Simulations of NSMs suggest multiple sites where element synthesis occurs (see \citet{fernandez2016EMSig} and \citet{Shibata2019Merger} for recent reviews).
Most models show an initial ejection of matter on dynamical timescales, due to some combination of tidal interactions during the NS inspiral, hydrodynamic \lq\lq squeezing" of material from the contact interface between coalescing NSs, and pulsations of a merged hyper- or supramassive NS remnant that can unbind additional material from the remnant's surface~\citep{Oechslin2007Relat,Bauswein2013SYSTEM,Hotokezaka2013Masse,Lehner2016Unequal}.
While tidally ejected outflows are traditionally expected to be neutron-rich enough to produce a main \rp{}, i.e. nuclei between $A\sim130$ and $A\sim195$ (often referred to as second and third \rp{} peak) and beyond~\citep{Goriely2011rPROC,Korobkin2012Onth,Mendoza-Temis2015Nuclea}, there is less consensus on the nucleosynthesis that takes place in the collisional and pulsational ejecta.
Some simulations of these components \citep{Goriely2011rPROC,Bovard2017r-proc,Radice2016Dynami} find a low electron fraction ($Y_{e}$) consistent with heavy \rp{} production.
Others, however, predict that weak current interactions can push the distribution of $Y_{e}$ toward higher values, in some cases raising it enough that nucleosynthesis is restricted to a light \rp{} that fails to reach the third peak~\citep{Oechslin2007Relat,Wanajo2014PRODUC,Sekiguchi2015Dynamic}.
In addition to dynamical ejection channels, the formation of an accretion disk~\citep{Surman2008r-Proc,Perego2014Neutri,Just2015Compre,Wu2016Produc,Siegel2017Three} and possible hypermassive NS~\citep{Baumgarte1999Onthe,Lippuner_hyper,vanPutten_hyper,Metzger2018Magnetar,ciolfi2020magnetically} after the merger is expected to drive further outflows.
These may be neutron rich enough to produce \rp{} material~\citep{Surman2006Nucleo}, but could also be responsible for material closer to the iron peak~\citep{Surman2005Neutri,Arcones2011PRODUC}.

While it is currently difficult to identify individual main \rp{} isotopes from analyses of observed spectra of GW170817, it is widely accepted that the light curve and spectral energy distribution of the associated transient are consistent with at least some portion of the outflows producing lanthanides~\citep{Kasen2017origi,Drout2017Light,Chornock2017TheEl,NichollTheEl}.
Evidence supporting this conclusion can be found in the luminosity and morphology of the light curve, and in particular by the uniquely red color of the emission, a signpost for \rp{} compositions~\citep{Barnes2013Effec, Tanaka2013Radia}. 
Evidence also suggesting a rapidly evolving blue component inconsistent with lanthanide enrichment points to the relevance of multi-component models. 
Specifically, these include a red, dim component which is rich in lanthanides and/or actinides~\citep{Kasen2017origi} and a blue component that is effectively free of these high opacity elements, which could be caused by material closer to the iron peak~\citep{Evans2017Swift, Miller2019Fullt}.
Models that account for the composition-dependent opacity of the \rp{} ejecta and the energy released by the decay of \rp\ elements can be used to estimate the amount and composition of \rp{} material ejected in the merger~\citep{Radice2018binar}. 
This is an important quantity to estimate because it is not clear if all the \rp{} material is made in compact object mergers, or if a second type of explosive event is required to account for all the \rp{} material in the galaxy~\citep{Travaglio2004Galact,Mennekens2014Massiv,Cote2018TheOr,Cote2019Neutro}.

Following the evolution of material made via the \rp{} requires tracking the production and decay of nuclei far from stability, where nuclear properties are often unmeasured (for review, see \citet{Mumpower2016Theim} and recent works \citet{Mumpower2015Theim,Lippuner2015r-pro,Barnes2016Radio,Zhu2018Califo,Wu2019Finger}).
Some light curve calculations have previously suggested the radioactive decay of the \rp{} material that powers the KN to be dominated by $\beta$-decay, and in fact, models assuming high-opacity lanthanide production seem to successfully interpret late-time observations with a power-law fitted $\beta$-decay heating rate~\citep{Waxman2019Latet}.
However, more detailed tracking takes into account different decay modes~\citep{Wanajo2018Physic,Hotokezaka_2020,Korobkin_2020,fujimoto_isomers}, which affect the energy released as well as the way that energy is converted to thermal photons, i.e. its thermalization efficiency.
Uncertainties in the ejecta composition and dominant decay modes determining energy deposition therefore propagate to uncertainties in the predicted light curve.
Since KN energy generation is therefore fundamentally sensitive to uncertainties in nuclear properties, these join other uncertainties such as the equation of state~\citep{Radice2018GW1708,Malik2018GW1708,Abbott2018GW1708,Coughlin2018Constr,Carson2019Equati,Gamba2019Theim}, neutrino transport~\citep{Ruffert1997Coales,Dessart2008NEUTRI,Abdikamalov2012ANEW,Fernandez2013Delaye,Richers2015MONTE} and flavor transformation~\citep{Malkus2012Neutri,Zhu2016Matte,Frensel2017Neutri,Wu2017Imprin,Richers2019Neutri}, and atomic opacity calculations~\citep{Fontes2015Relati,Tanaka2019System} as areas that merit closer inspection. Evaluating the effects of nuclear processes with other astrophysical and thermodynamic data from theoretical simulations allows for more informed models to aid in the interpretation of future KN signals. To move forward quantitative descriptions for the uncertainty range nuclear inputs can generate, here we explore a wider range of nuclear mass models and fission properties than have been considered by previous works \citep{Metzger2010Electr,Wanajo2018Physic,Barnes2016Radio,Wu2019Finger,Radice2018binar,Lippuner2015r-pro,Vassh2019Using,giuliani2019fission,Eichler2019Probing,Even2019Compos,Hotokezaka_2020,Korobkin2020Axisymm}.

In this paper, we explore the impact of variation in nuclear physics inputs for a range of initial astrophysical conditions on the  energy generation and radioactively-powered emission from lanthanide-rich outflows believed to be responsible for the red components of kilonovae.  
While the GW170817 kilonova has been extensively studied there are still uncertainties which remain; for example detailed numerical simulations and semi-empirical models have successfully explained some, but not all of the GW170817 kilonova evolution, e.g. \citet{Chornock2017TheEl}.
Furthermore, many models require two or three substantial high-mass outflow components which is in tension with numerical relativity simulations \citep{Villar2017TheCo}. 
Thus, it is worth comparing the results of nucleosynthesis-informed predictions directly with the observed light curve before additional manipulation of these results occurs. 
We anticipate that our results, which emphasize contributions to the light curve from species with $A \gtrsim 120$, will be useful not only for elucidating aspects of symmetric mass NSMs, such as GW170817, but also asymmetric NSMs and black hole-neutron star (BHNS) mergers, which are suggested by simulation to expel a greater amount of very neutron-rich ejecta than the symmetric mass case.

This paper is structured as follows. We introduce the astrophysical conditions and nuclear inputs used in our nucleosynthesis simulations, as well as outline the radiation transfer methods we use for our models in Section~\ref{sec:method}. 
In Section~\ref{sec:source}, we then evaluate the scale of uncertainties in ejecta composition and nuclear heating in order to provide a comprehensive estimate of their effect on nucleosynthesis-informed light-curve modeling in Section~\ref{uncertainties}.
Finally, in Section~\ref{fission}, we identify key nuclei and reactions that drive the KN heating.

\section{Method}\label{sec:method}
In this section we describe our simulation pipeline, which begins with the choice of astrophysical conditions; we choose a wind-type model with a variety of initial electron fractions as a proxy for the full possible range of hydrodynamic and thermodynamic outflow conditions in a NSM. 
We use each of these thermodynamic/hydrodynamic trajectories in a separate nuclear reaction network calculation for thirty-two different combinations of nuclear inputs. 
The output of these calculations are used in a semi-analytic KN model that estimates the system's emerging luminosity.
We include in this model approximations to the full opacity calculations, anticipating that future measurements and theoretical calculations of opacities will contribute to the ultimate goal of exact predictions of KN light curves.  
Our model provides a picture of the general trends of a lanthanide rich component of KN light curves and estimates for light curve uncertainties through a survey of nuclear inputs in a variety of conditions.

\subsection{Astrophysical conditions}\label{method:astro}
The astrophysical parameters that describe the initial conditions of \rp{} outflows launched by compact object mergers giving rise to KNe have been studied extensively, for example in \citet{Korobkin2012Onthe, Martin2015Neutri, Lippuner2015r-pro,Wanajo2018Physic,miller2019Fulltr}. 
The nuclei synthesized in NSM ejecta are determined by a number of factors, including the entropy, the gas' rate of expansion, and initial neutron richness of the ejecta. 
Here we choose astrophysical conditions typical of merger accretion disk winds and use variations in the initial electron fraction, $Y_e$, of the ejecta as a proxy for all astrophysical variations.

The base astrophysical trajectory is a standard parameterized wind \citep{Panov+2009}, with initial entropy per baryon in units of Boltzmann constant of $s/k=40$ and an expansion timescale of $20 \, {\rm ms}$, conditions similar to those found in the multidimensional NSM simulations of, e.g., \citet{Just2015Compre,Radice2018binar} and as applied in \citet{Zhu2018Califo}. 
Note that nucleosynthesis proceeds slightly differently at different entropy, with a higher entropy working similarly to a lower $Y_e$, as both can increase the \rp{} reach \citep{Meyer_1997}. 
Simulations begin in nuclear statistical equilibrium at a temperature of  $T = 10$ GK with seed nuclear abundances generated using the SFHo equation of state ~\citep{Steiner2012Core-c}. 
For all simulations, we use the same evolution of density as a function of time (expansion rate) but choose a variety of initial electron fractions. 
With different initial electron fractions, nucleosynthesis processes and their corresponding energy generation as a function of time are different.  
Therefore, the temperature of each simulation must be determined separately; and we use an adiabatic expansion law which is modified assuming a $10\%$ nuclear reheating efficiency as in~\citet{Holmbeck2018Actini,Vassh2019Using}.
As the energy generation is different for each set of nuclear inputs, each simulation we consider has a unique thermodynamic history.   

Based on an initial survey of $Y_e$ from  $0.01$ to $0.30$ (with increments of 0.01) we find that a subset of these is sufficient to capture most of the effects on the light curve. 
Therefore, to keep simulation expense low without losing a reasonable coverage of electron fractions, we use $Y_e$ of $0.02$, $0.12$, $0.14$, $0.16$, $0.18$, $0.21$, $0.24$, and $0.28$ in our nucleosynthesis calculations, which we describe next.

\subsection{Nucleosynthesis}\label{method:nucleosyn}
For the nucleosynthesis simulations, we use the Portable Routines for Integrated nucleoSynthesis Modeling (PRISM)~\citep{Sprouse2019Inprep} reaction network developed at the University of Notre Dame and Los Alamos National Laboratory. 
PRISM is designed to work with prepared input files of astrophysical conditions, nuclear mass models, and nuclear reaction channels (such as charged particle reactions, neutron capture, photodissociation, $\beta$-decay, $\beta$-delayed neutron emission, neutron-induced fission, $\beta$-delayed fission, and spontaneous fission as in~\citet{Zhu2018Califo,Vassh2019Using}).
For experimentally measured nuclei, we use measured masses from AME2016 from~\citet{Wang2017TheAM} and experimental data from NUBASE2016 from~\citet{Audi2017TheNU}, including $\alpha$-decay, $\beta$-decay, electron capture, neutron emission, and spontaneous fission data.  
Since we are interested in uncertainties from theoretical nuclear inputs for nuclei heavier than iron, we use the JINA Reaclib nuclear reaction database \citep{Cyburt2010THEJI} for charged-particle  and light-nuclei reactions, but otherwise explore the effect of a wide range of predictions for as yet unmeasured masses and reactions rates for heavier nuclei. 

Where experimental data is not available, we use nuclear physics inputs starting from the eight theoretical nuclear models listed in Table~\ref{tabmasstable}, and all the reaction rates are determined consistently with the associated masses \citep{Mumpower2015}. 
The neutron capture and neutron-induced fission rates are recalculated with each chosen mass model using Los Alamos National Laboratory statistical Hauser-Feshbach code CoH~\citep{Kawano2016Statis}.
The $\beta$-decay strength functions are from~\citet{Moller2019Nuclea},  and the relative probabilities for $\beta$-decay, $\beta$-delayed neutron emission, and $\beta$-delayed fission are calculated using the QRPA+HF framework \citep{Mumpower2016Neutro,Mumpower2018betad}.
Theoretical $\alpha$-decay rates are obtained with a Viola-Seaborg relation using Q-values calculated from each chosen mass model and parameters fit to known data~\citep{Vassh2019Using}.
For neutron-induced fission, $\beta$-delayed fission, and spontaneous fission, we use either a symmetric ($A_{fragment}=\frac{1}{2}A_{fission}$) or a double Gaussian fragment distribution from~\citet{Kodama1975R-proc}; most modern fission yield predictions fall somewhere between these extremes. 
We make an exception for the fission yields for \Cf, in which case we use the yield distribution described by \citet{Zhu2018Califo}.
For spontaneous fission half-lives, we apply a barrier-height-dependent prescription from~\citet{Karpov2012DECAY,Zagrebaev2011Produc} (KZ) or the parameterization from~\citet{Xu2005System} (XR).  
Fission rates are updated with the fission barrier predictions most closely associated with a given mass model as in~\citet{Vassh2019Using}.
That is, for the ETFSI model we apply ETFSI barriers, for HFB-22 and HFB-27 we apply HFB-14 barriers, for the TF model we apply TF barriers, and for all other cases, including FRDM2012, we apply FRLDM barriers.

\startlongtable
\begin{deluxetable}{l c r}
\tablewidth{0pt}
\tablecaption{\label{tabmasstable}Mass Models Used in the Nucleosynthesis Simulations.}
\tablehead{\colhead{Mass Model} &  \colhead{Abbreviation} & \colhead{Masses, Fission Barriers}}
\startdata
Finite-Range Droplet Model & FRDM2012 & ~\citet{Moller2016Nuclea,MollerBH2}\\
Duflo and Zuker  & DZ33 & ~\citet{Duflo1995Micros, MollerBH2}\\
Hartree-Fock-Bogoliubov 22 & HFB22 & \citet{HFB22,HFBBH}\\
Hartree-Fock-Bogoliubov 27 & HFB27 & \citet{Goriely2013Hartre,HFBBH}\\
Skyrme-HFB with UNEDF1 & UNEDF1 & \citet{Kortelainen2012Nuclea,MollerBH2}\\
Skyrme-HFB with SLY4 & SLY4 &\citet{Chabanat1998ASkyr,MollerBH2}\\
Thomas-Fermi &  TF & \citet{TFmass,TFBH}\\
Extended Thomas-Fermi plus Strutinsky Integral &  ETFSI & ~\citet{Aboussir1995Nuclea,Mamdouh01}\\
Weiz\"{a}cker-Skyrme (WS3) &WS3& \citet{Liu2011Furthe,MollerBH2}\\
\enddata
\end{deluxetable}

\subsection{Thermalization efficiencies}\label{method:thermalization}
The total heat supplied to the light curve is a function of the energy emitted by radioactive decay and the efficiency with which that energy thermalizes, i.e., is converted into the thermal energy that ultimately forms the light curve.
The distribution of radioactively decaying nuclei evolves rapidly in the days and weeks following the merger. 
We keep track of the energy emitted through each channel ($\beta$-decay, $\alpha$-decay and fission) as a function of time for each nucleosynthesis simulation. 
This is important because there is not a one-to-one correspondence between a final abundance pattern and the history of heating and nuclear decays~\citep{Kawaguchi2019Divers}.

Different decay channels have different overall levels of thermalization efficiency.
Energy released in $\beta$-decays heats the ejecta less effectively than that released in $\alpha$-decay, which in turn is less efficient than fission, as shown in~\cite{Barnes2016Radio}.
This is partly because a substantial fraction of $\beta$-decay  energy is released as neutrinos, which free-stream out of the diffuse KN ejecta without interacting.
In contrast, in $\alpha$-decays and fission, most of the energy goes to massive particles, which thermalize more effectively in the ejecta. To capture these effects, we calculate the rate at which energy is produced by $\beta$-decay, $\alpha$-decay, and fission as a function of time.
We then apply analytic thermalization efficiencies from ~\citet{Kasen2019Radioa} (hereafter referred to as KB19) for each channel to estimate the total rate at which energy is thermalized, $\dot{Q}(t)$, as a function of time,
\begin{equation}\label{eq:totalheat}
\dot { Q } ( t ) = \sum _ { i } f _{ i } ( M_{\rm ej}, v_{\rm ej}, t ) \; \dot { q } _ { i } ( t ) \; M_{\rm ej}.
\end{equation}
In the above, $f_{ i }$ is the thermalization efficiency for the reaction channel $i$, which depends on the mass, $M_{\rm ej}$, and characteristic velocity, $v_{\rm ej}$, of the ejecta in addition to time, and $\dot{q}_{i}(t)$ is the nuclear heating released by channel $i$ per unit mass of ejecta, as determined from the reaction rates and Q-values of each nuclear reaction in each nucleosynthesis simulation.
We determine the thermalization efficiencies, $f_i(t)$, following the procedure outlined in \citet{Kasen2019Radioa} in which the thermalization of massive particles is assumed to have the form $f=(1+\tau)^{-n}$. 
Here, $\tau$ is the time scaled to a characteristic \lq\lq thermalization time", $t_{\rm th}$, at which thermalization starts to become inefficient.
Both $t_{\rm th}$ and the power-law index, $n$, depend on the parameters of the ejecta, the energetics of the decaying nuclei, the magnitude of the energy-loss cross-sections, and their dependence on particle energy.

\begin{figure}[!ht]
\fig{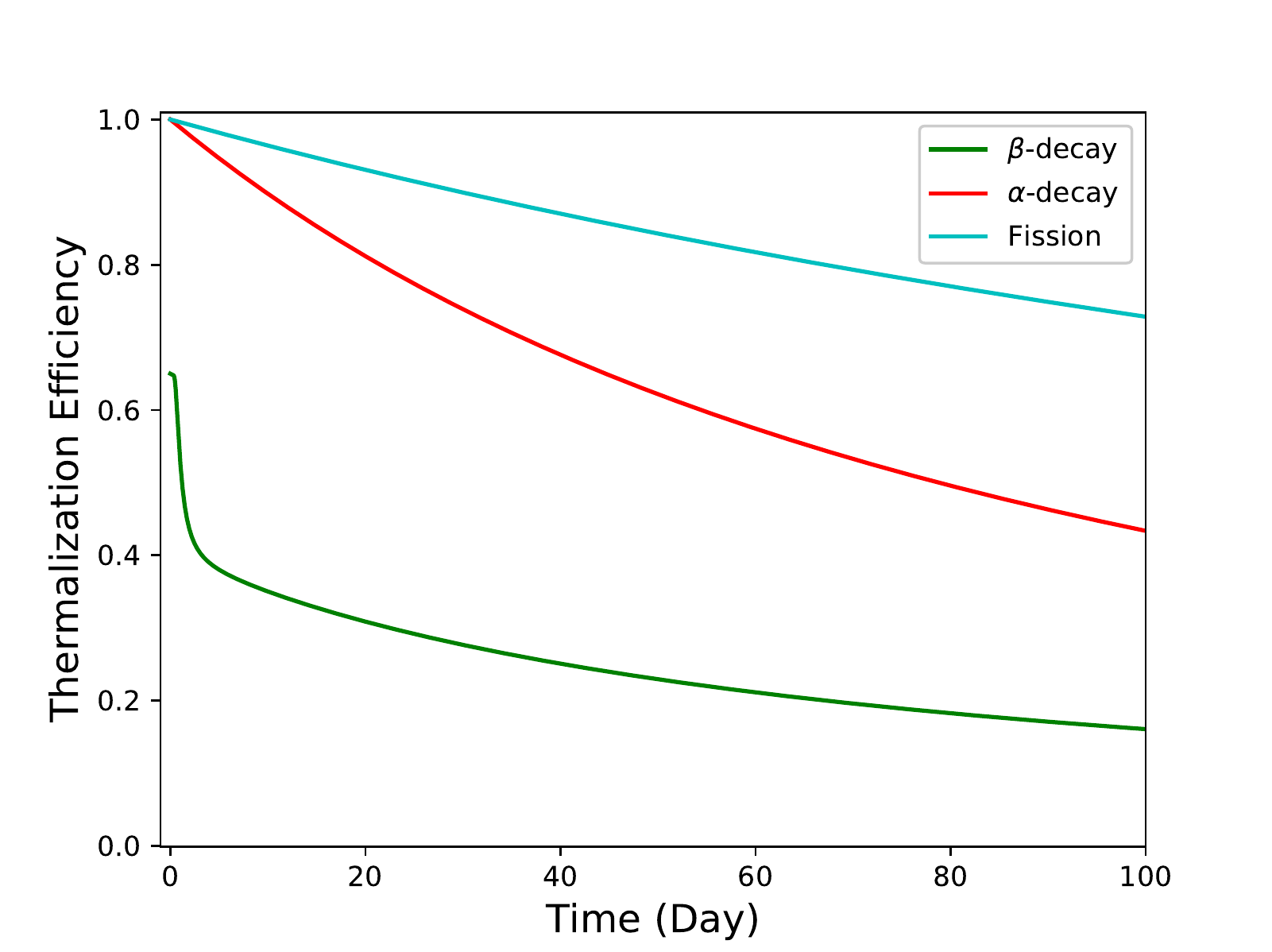}{0.65\textwidth}{}
\caption{\label{fig:thermalization}Thermalization efficiencies for $\alpha$-decay (red line), $\beta$-decay (green line), fission (cyan line) as described in Section~\ref{method:thermalization} using ejecta mass $M_{\rm ej}=0.05 M_\odot$ and the characteristic ejecta velocity $v_{\rm ej}=0.15c$.}
\end{figure}

The total thermalization efficiency for $\beta$-decay depends on the efficiencies for electrons and $\gamma$-rays, as well as the fraction of energy lost to neutrinos.  
We assume that 25\% of the energy goes to electrons, 25\% to $\gamma$-rays, and the remainder to neutrinos \cite{Barnes2016Radio,Hotokezaka2016Radioa}, which is consistent with the decays that are occurring in the calculations presented in the next sections. The electron efficiency follows the pattern defined above, with $n=1$, and $t_{\rm th}$ defined by (KB19 eqn. 43):
\begin{align}
t_{th,\beta} &= 12.9 \;  M_{0.01}^{2/3} \; v_{0.2}^{-2} \; \zeta^{2/3} \text{ days},
\label{eq:tth_b}
\end{align}
where $M_{0.01}$ is the ejecta mass in units of 0.01$M_\odot$, $v_{0.2} = v_{\rm ej}/0.2c$, and $\zeta$ is a constant defined to be close to unity (we adopt $\zeta=1$).  
For $\gamma$-rays, the thermalization efficiency is assumed to be comparable to the probability of absorption or scattering in the ejecta (KB19 eqns. 48, 53),
\begin{align}
f_{\gamma}(t) &= 1-\exp[t_\gamma^2/t^2], \\
  t_\gamma &= 0.3 \; M_{0.01}^{1/2} \; v_{0.2}^{-1}.
  \label{eq:tth_g}
\end{align}
The efficiencies for $\alpha$-decay and fission are the efficiencies of the corresponding particles.
For $\alpha$-decay,
\begin{align} 
f_{\alpha}(t) &= (1 + t/t_{\rm th,\alpha})^{-1.5}, \\
t_{\rm th,\alpha} &= 2 \; t_{\rm th,\beta},
\end{align}
while for fission fragments,
\begin{align}
    f_{\rm f}(t) &= (1+t/t_{\rm th,f})^{-1}, \\
    t_{\rm th,f} &= 4 \; t_{\rm th,\beta}.
\end{align}
The scalings and choice of $n$ come from comparing the typical energies and energy-loss rates of the different particle types~\citep{Barnes2016Radio}.
The thermalization efficiencies $f_i(t)$ for each decay channel are plotted in Fig.~\ref{fig:thermalization}.

\subsection{Opacity}\label{method:opacity}
We estimate the opacity of the KN ejecta from the composition of each simulation at a representative time $t_\kappa=1$~day after the merger.
The assumption of constant composition is standard in radiation transport models of KNe \citep[][among many others]{Kasen2017origi,Tanaka2019System,Even2019Compos}, and is justified by the much slower evolution of the composition on these longer timescales relative to the first minutes and hours after free neutron depletion.
In particular, the mass fractions of the highest opacity species are not found to change dramatically over the course of the KN, and the effects of temperature are expected in this epoch to have a greater effect on the opacity than any lingering compositional drift \citep{Kasen2013Opaci,Tanaka2019System}.

Opacity has been shown to depend critically on bound-bound structure, and because we lack atomic data for all atomic species, we carry out our opacity calculation for a carefully constructed modified composition. 
We take the mass fraction in all s- and p- block elements and, for the purpose of calculating opacity only, assume that this mass resides in calcium ($Z=20$).
Since the opacities from d- and f-block dominate the total opacity of the gas, we modify the full composition in a manner that preserves the total mass fractions in d- and f-blocks. 
In the d-block, we evenly divide the mass fractions in d-block elements among $Z=21-28$.  
In the f-block (lanthanides and actinides) we take care to preserve the mass fraction in each column of the periodic table, since atomic opacities vary both among and within the blocks of the periodic table~\citep{Kasen2013Opaci,Tanaka2019System}, and because f-block abundances have the strongest impact on the total opacity.
In our modified composition, lanthanide and actinide mass fractions are distributed as described above among $Z=59-70$.  

We calculate the Planck mean of the expansion opacity \citep{Karp.ea.1977_line.exp.opac} for the modified compositions described above at a range of temperatures, assuming local thermodynamic equilibrium and adopting a mass density $\rho=10^{-14}$~cm$^2$~g$^{-1}$, typical of conditions in the ejecta on KN timescales.
The results of these calculations inform the 
temperature-dependent opacity function we use to compute semi-analytic light curves:
\begin{equation}
\kappa =
\begin{cases}
\kappa_{\rm max} \left(\frac{T}{4000 \; \rm{K}}\right)^{5.5}, \: &\text{ $T < 4000$ K } \\
\kappa_{\rm max} \: &\text{ otherwise }
\end{cases}
\end{equation}
In the equations above, $\kappa_{\rm max}$ is the maximum value from the Planck mean opacity calculation, and is unique for each composition.
The steep decline in $\kappa$ for $T < 4000$ K accounts for the opacity lost as the higher-energy electronic states of high-opacity lanthanide and actinide atoms and ions are depopulated due to the decreasing amount of available thermal energy in a cooling gas.

While opacity does have an effect on our calculations, since we are looking at lanthanide-rich scenarios, it tends to be less significant than other uncertainties such as the nuclear heating. We explore the effects of different lanthanide and actinide mass fractions, and therefore changing opacity, in Section \ref{sec:curves}.

\subsection{Light curve}\label{method:lc}
The effective heating, $\dot Q(t)$ and opacity, $\kappa(T(t))$ are input into a semi-analytic light curve model based on that of \citet{Metzger2017kilon}, in which the KN ejecta is discretized as a set of concentric shells of mass $M_{v}$.
We choose a power-law mass density profile  
\begin{equation}
\rho(v,t) = \left( \frac{3}{4 \pi} \right )
\frac{M_{ej}}{v^3 t^3}
\left( \frac{v}{v_0} \right ) ^{-3},
\end{equation}
where $v$ is the velocity coordinate, which extends from $v_0 = 0.1c$ to $v_{\rm max} = 0.4c$.
Since we are interested in the red KN component containing the most massive \rp\ nuclei, we adopt as our fiducial model an ejecta with mass $M_{\rm ej} = 0.05 M_\odot$ and $v_{\rm ej} = 0.15c$, typical of the values inferred for the outflow powering the red emission of the KN associated with GW170817~\citep{Kasen2017origi,Kasliwal2017Illumi,Tanaka2017Kilono,Drout2017Light,Perego2017at2017}. The effect of varying ejecta mass is explored in Section~\ref{uncertainties}. 
The characteristic ejecta velocity, defined in terms of the ejecta's total kinetic energy, $E_{\rm kin}$, through $v_{\rm ej} = (2 E_{\rm kin}/M_{\rm ej})^{1/2}$, is used to calculate thermalization efficiencies as expressed in Eq.s~\ref{eq:tth_b} and~\ref{eq:tth_g}.
In our fiducial model, $v_{\rm ej} = 0.15c$.
We consider 100 shells and evolve the thermal energy, $E_{v}$, of a shell of mass $M_{v}$ at velocity $v$ as 
\begin{equation}
    \frac{{\rm d} E_{v}}{{\rm d} t } = \frac{M_{v}}{M_{ej}} \dot{Q}(t, v)-\frac{E_{v}}{t} - L_{v},
\end{equation}
where the first term on the right hand side is the effective heating from (thermalized) radioactivity, the second term accounts for loss of radiation energy due to adiabatic expansion, and the final term represents the outgoing luminosity from the mass shell. 
The luminosity comprises photons that diffuse and free-stream out of the shell, and is approximated as
\begin{equation}
    L_{v} = \frac{E_{v}}{t_{\rm d,v} + t_{\rm lc}}.
\end{equation}
Here, $t_{\rm lc}$ is the light-crossing time and $t_{\rm d,v}$ the diffusion timescale at velocity $v$.
These quantities are given by
\begin{align}
    t_{\rm lc} &= vt/c, \: \text{ and} \\
    t_{\rm d,v} &= {M_{>v} \kappa(T_v(t))}{4 \pi vt c }
    \label{eq:tdiff}
\end{align}
In Eq.~\ref{eq:tdiff}, $M_{>v}$ is the amount of mass exterior to the mass shell with velocity $v$, i.e. the amount of mass which has velocity greater than $v$. 
The opacity of a given layer, as described above, depends on the temperature, $T_v$, which we calculate from $E_{v}$ assuming the ejecta is radiation dominated. 
The total luminosity at time $t$ is the sum of the luminosity escaping from each layer. 

\section{Uncertainties Stemming from  Nuclear Physics Inputs}\label{sec:source}  

In this section, we examine the effects of theoretical model uncertainties on both lanthanide and actinide mass fractions as well as the effective heating rate, using the methods described in Sections~\ref{method:nucleosyn} and~\ref{method:thermalization}. 
The composition plays a key role in determining the opacity of the ejecta and, together with the effective heating rate, is a critical factor influencing the magnitude and shape of the light curve. 
\begin{figure}[h!]
  \gridline{
  \fig{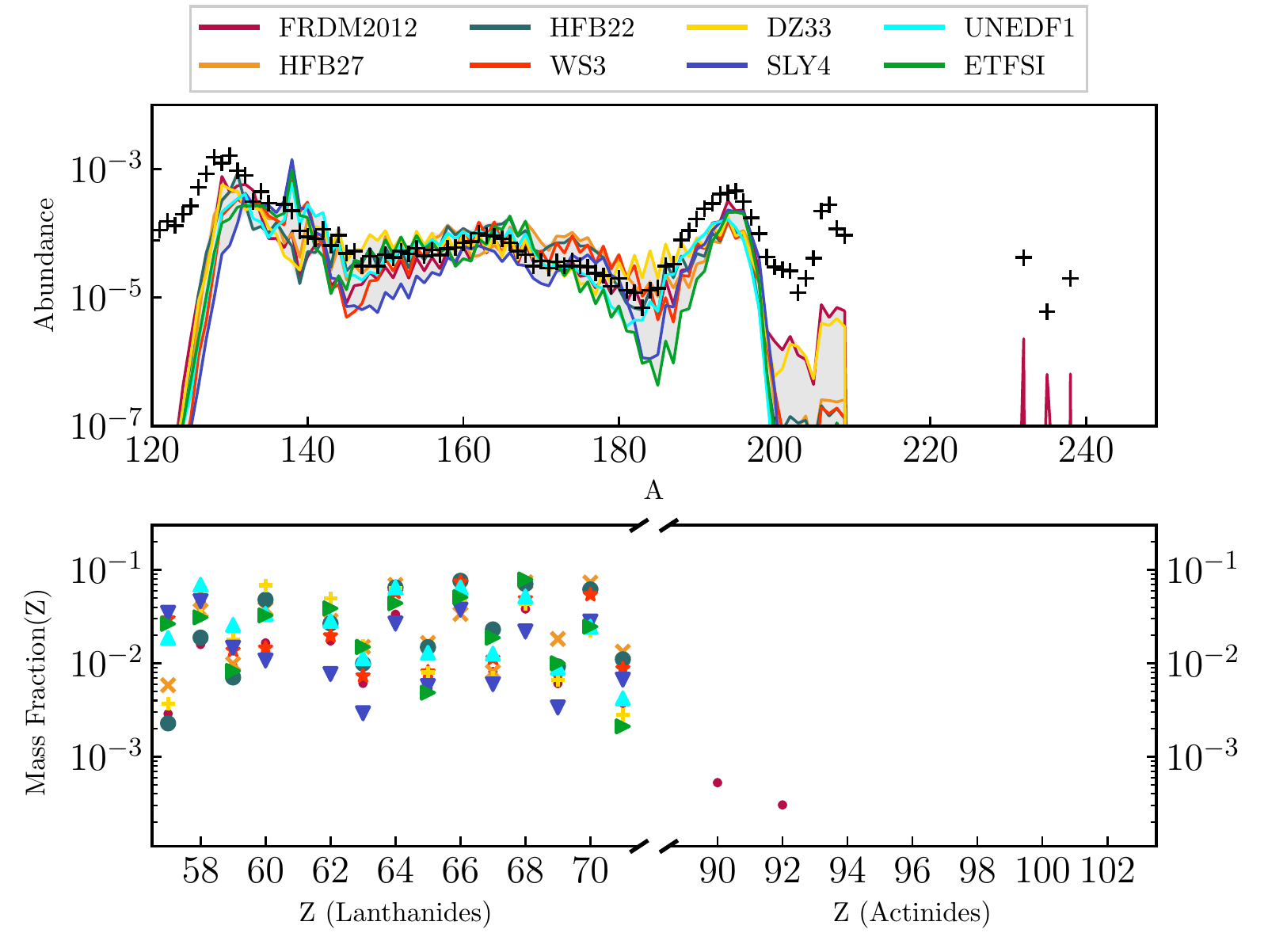}{0.49\textwidth}{(a) Simulations with $Y_e=0.24$.}
  \fig{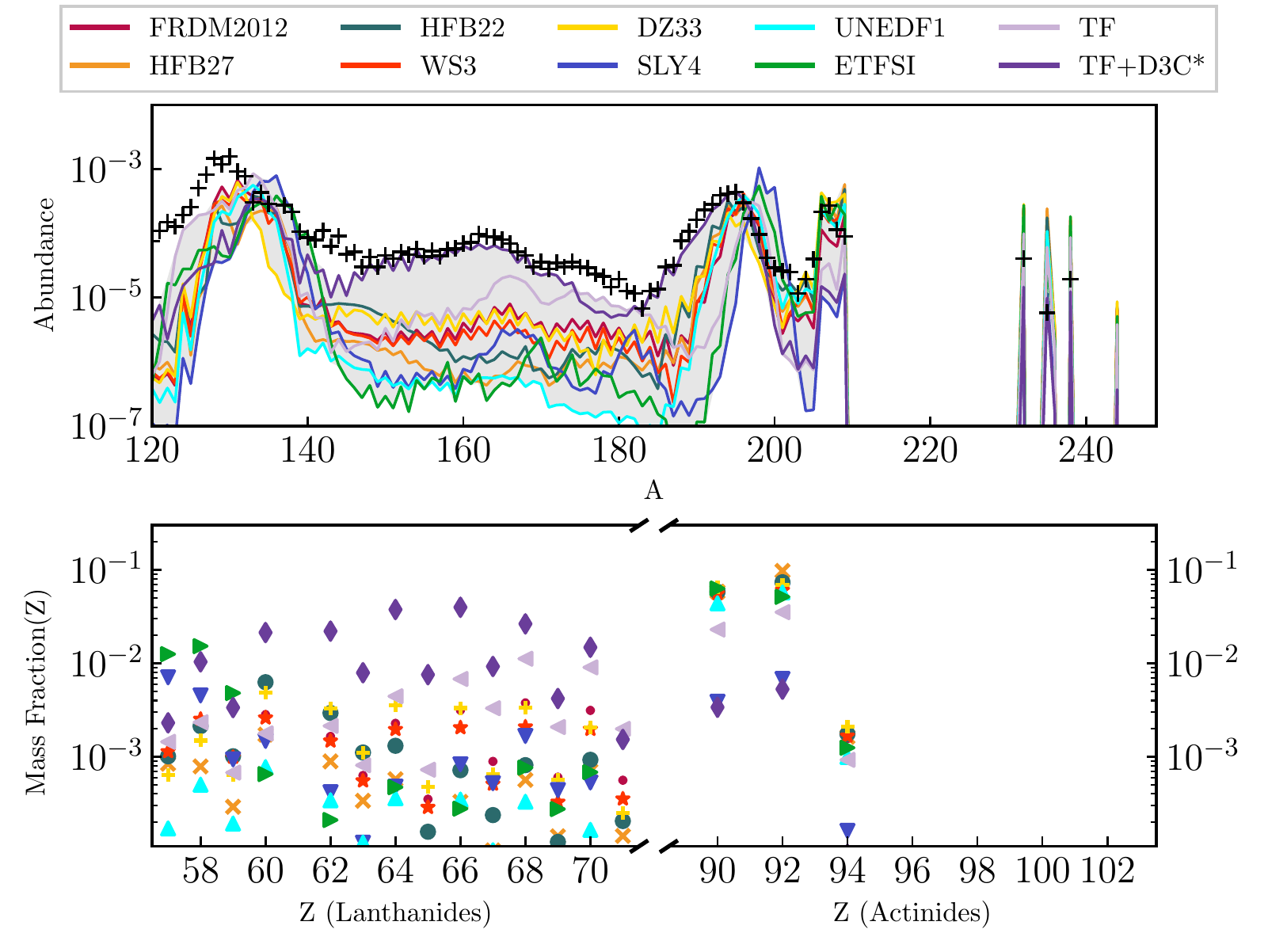}{0.49\textwidth}{(b) Simulations with $Y_e=0.16$.}}
  \caption{\label{fig:abye} Abundance pattern (top panel) and mass fraction of lanthanide and actinide elements (bottom panel) at two different electron fractions ($Y_e$).
  All simulations in these figures use symmetric fission yields and Karpov (KZ) spontaneous fission rates. In the top panels, black plus markers represent the solar \rp{} abundance~\citep{Sneden2008Neutro}. In all plots the solar data is scaled by comparing the average abundance in the third peak to the value obtained given FRDM2012 based nuclear inputs, symmetric fission yields, and KZ spontaneous fission rates.}
\end{figure}
We begin with an exploration of the effects of nuclear inputs, keeping in mind that the degree to which this uncertainty comes into play depends on the astrophysical conditions. 
For illustrative purposes, we consider simulations with two different initial electron fractions of $Y_e=0.24$ and $Y_e=0.16$; the former has minimal fission and the latter has substantial fission.
For each calculation, we vary the nuclear inputs using the combinations of mass models and fission barriers previously described,
with masses propagated to neutron capture and $\beta$-decay rates as described in Section \ref{sec:method}, in order to explore the scale of uncertainties in effective heating and final abundance patterns.

We show the final abundance and mass fractions of lanthanides and actinides of simulations with the aforementioned initial electron fractions in Fig.~\ref{fig:abye}.
In the lower panels of Fig.~\ref{fig:abye}, we see that both $Y_e=0.24$ and $Y_e=0.16$ have a sum of lanthanide and actinide mass fractions larger than $10^{-3}$, which is generally considered to be a lower threshold for a red KN \citep{Chornock2017TheEl}, as well as larger than $10^{-1.5}$, which is a lower limit for \rp~enhanced metal-poor stars \citep{Ji_2019}. 
High electron fraction ($Y_{e}=0.24$) scenarios show more significant lanthanide production than actinide production when compared with lower electron fraction ($Y_{e}=0.16$) scenarios.
The difference can be attributed primarily to the different numbers of neutrons available for capture in the two scenarios.
The upper panels of Fig.~\ref{fig:abye} show the final abundances (at $\sim0.3$ Gyr) of \rp{}~elements with different nuclear inputs, with the grey band indicating the spread. 
Since the simulations with $Y_e=0.16$ access fissioning isotopes, we include two additional nuclear models to show cases which find a relatively low population of actinide species at late times, TF and TF+D3C$^{*}$. 
This behavior is due in part to the relatively low fission barriers of the TF model near $N=184$, which increases the participation of neutron-induced and $\beta$-delayed fission. 
Species are cleared out of the actinide region even more when TF is coupled with the $\beta$-decay prescription of D3C$^{*}$ \citep{Marketin2016Beta}, which predicts $\beta$-decay rates faster than those calculated based on \cite{Moller2019Nuclea} above the $N=126$ shell closure, quickly decaying material into fissioning regions during times when neutron-induced fission is most active.
Simulations with $Y_e=0.24$ have a path that is significantly closer to stability, where the discrepancies between the masses predicted by the different models are more modest, which explains the smaller spread in simulated abundances and mass fractions.
Each of our calculations begins with a determined initial abundance calculated based on the nuclear model and $Y_e$. 
Thus, for both scenarios, the spread shown in this figure is an underestimate of the true uncertainty due to nuclear inputs, since uncertainties in the charged particle reactions that create the initial abundance of each trajectory were not taken into account in our analysis.
\begin{figure}[h!]
  \gridline{
  \fig{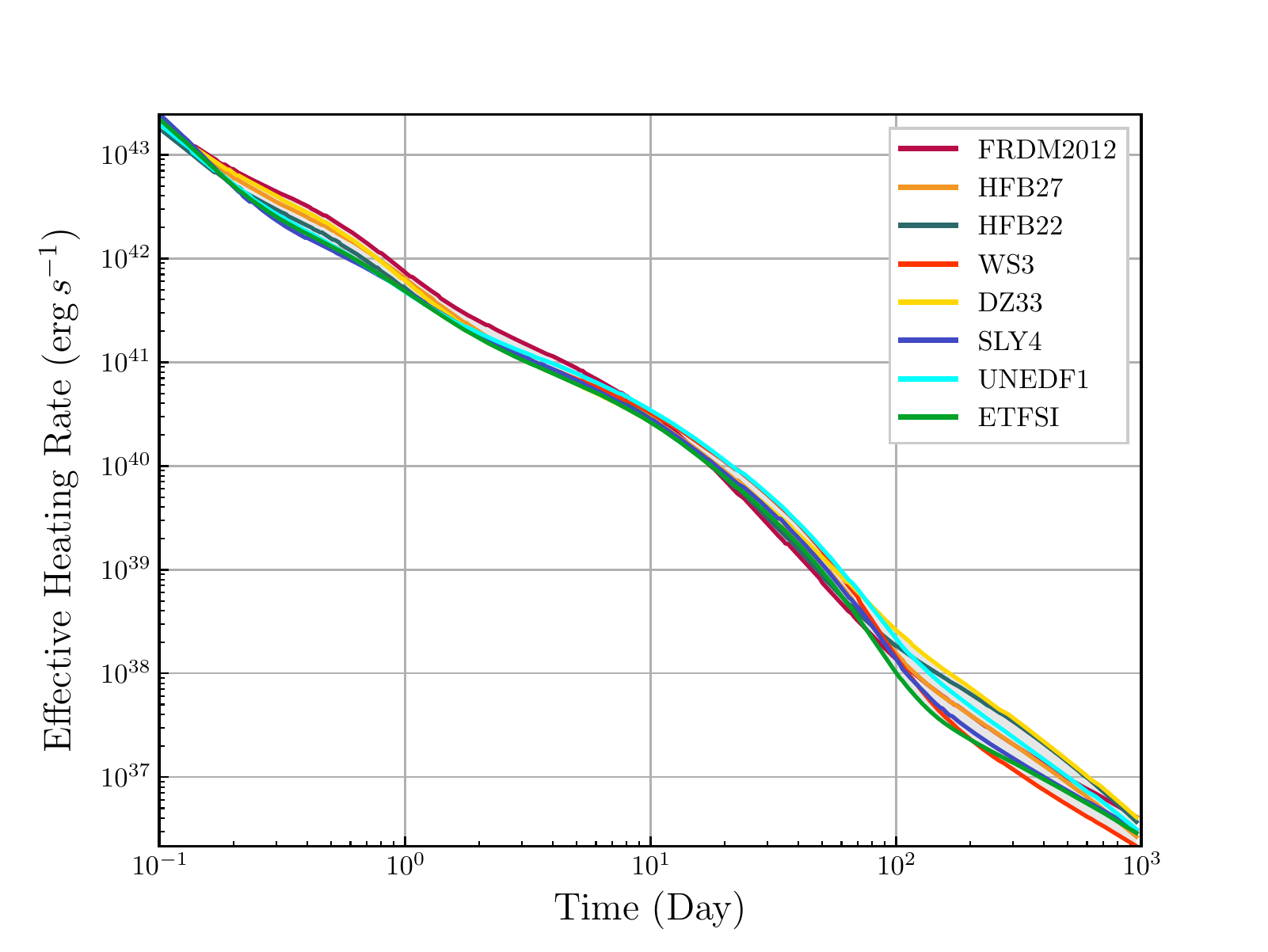}{0.49\textwidth}{(a) Simulations with $Y_e=0.24$.}
  \fig{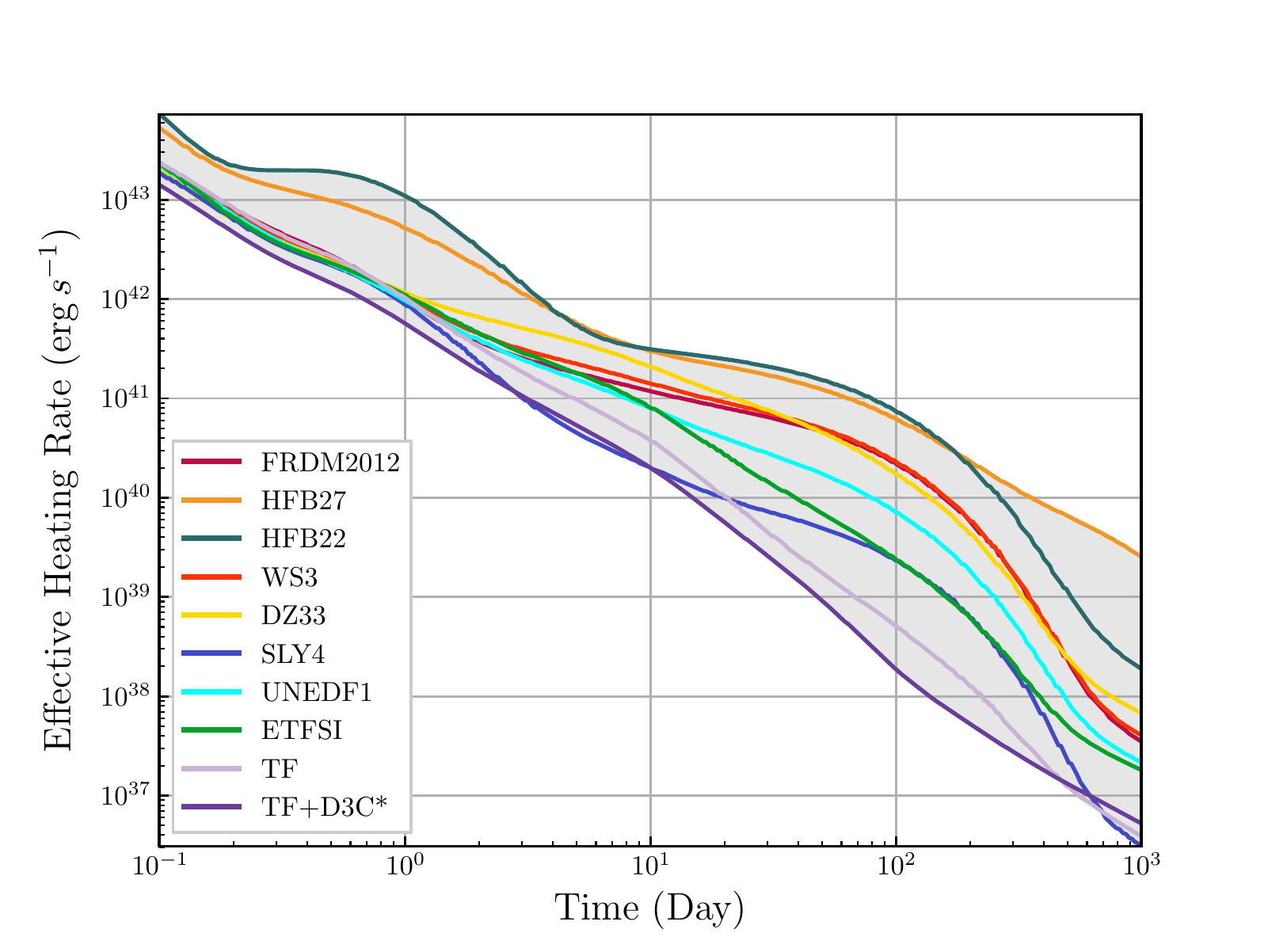}{0.49\textwidth}{(b) Simulations with $Y_e=0.16$.}
  }
  \caption{\label{fig:heatyesk}Comparison of total effective heating rates are shown for the same set of models as in Fig.~\ref{fig:heatyesk}; all models assume $M_{ej} = 0.05$ and $v_{ej}= 0.15$.  The effective heating rate for an individual nuclear model is shown as a colored line. 
  The grey band shows the range of effective heating rates from all simulations.}
  \end{figure}
\begin{figure}
    \gridline{
    \fig{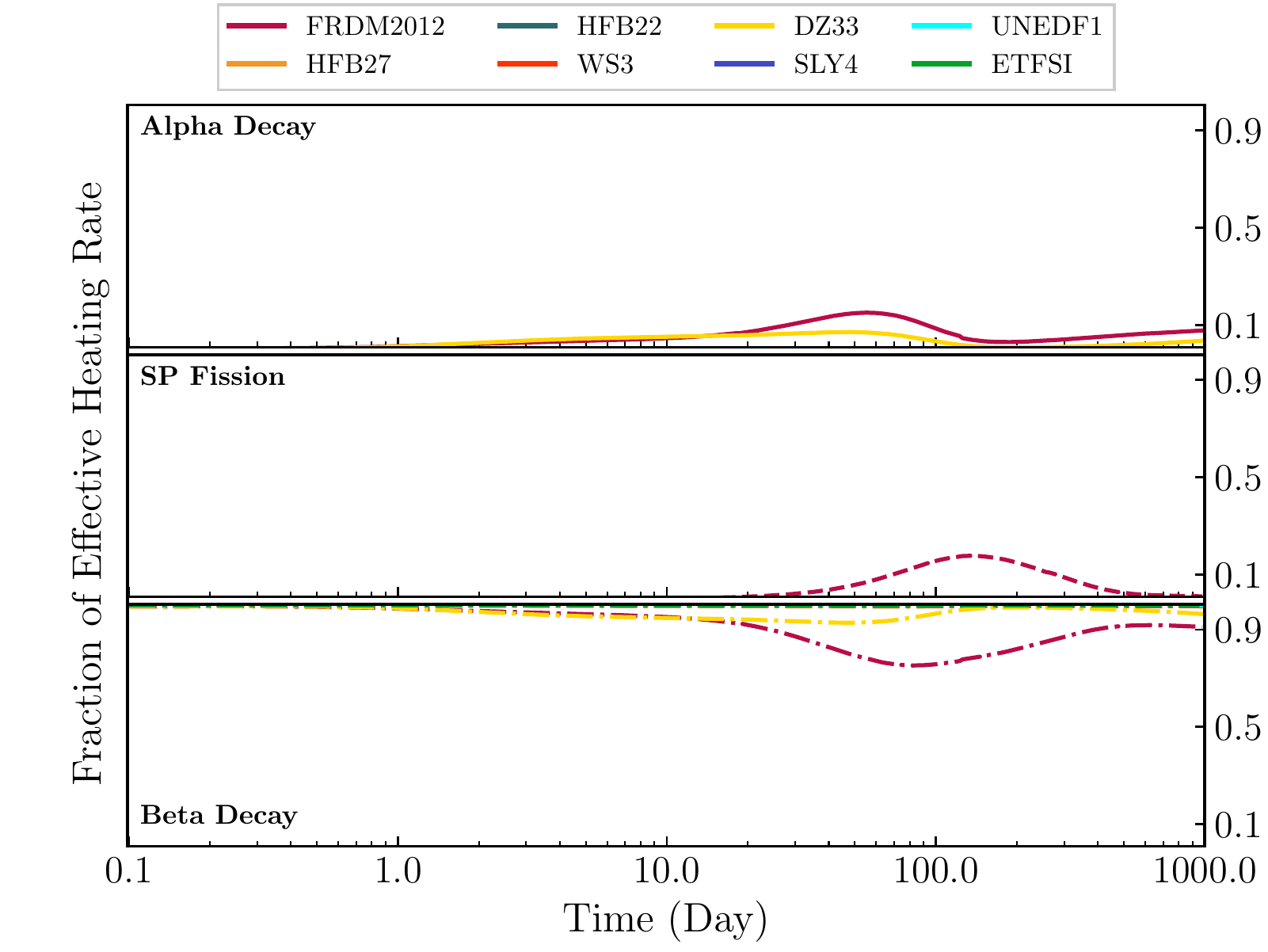}{0.49\textwidth}{(a) $Y_e=0.24$}
    \fig{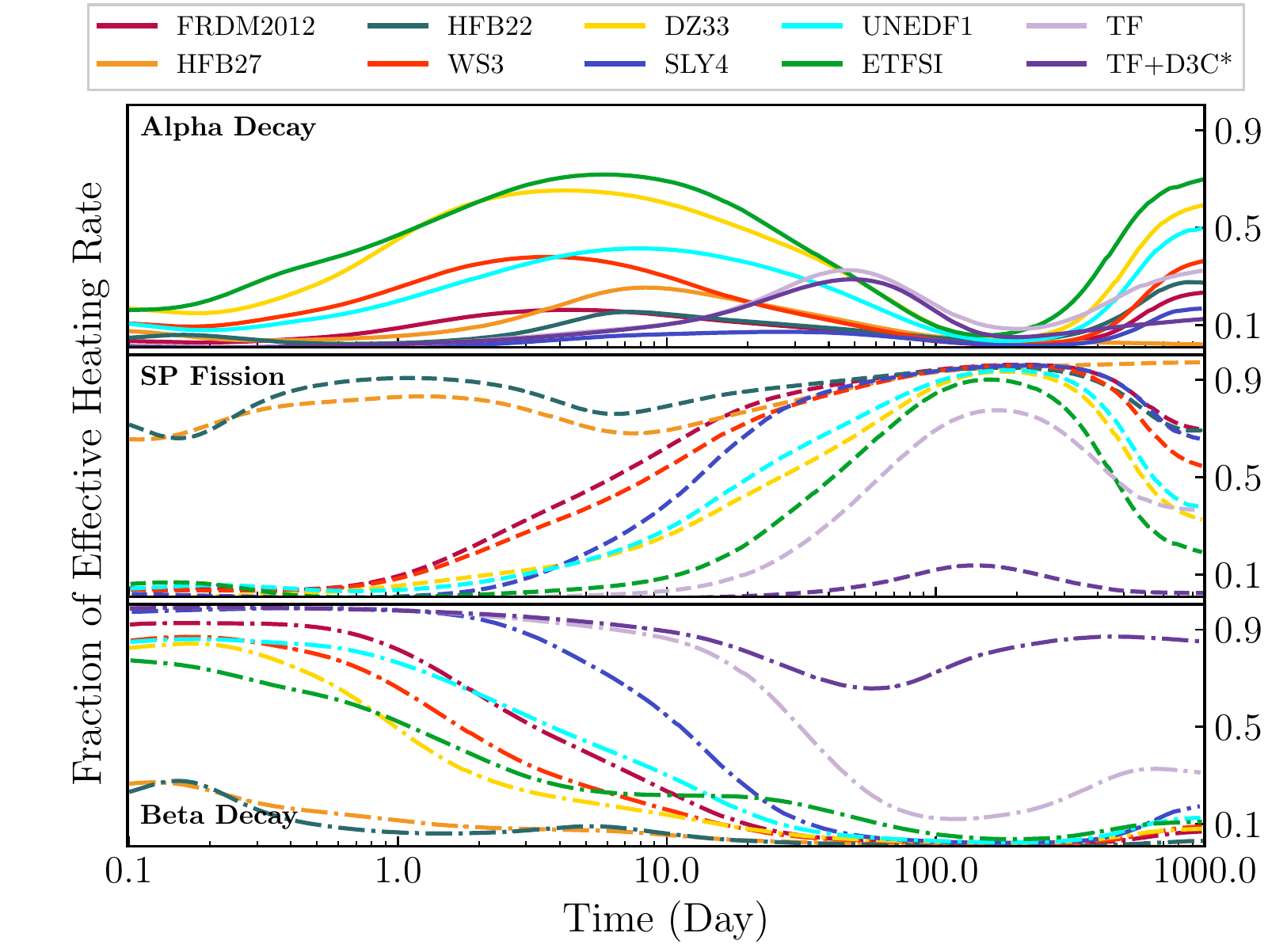}{0.49\textwidth}{(b) $Y_e=0.16$ }             
    }
    \caption{\label{fig:fractionss1624}Fractional heating rates assuming different theoretical nuclear models. All simulations are performed with symmetric fission yields and KZ spontaneous fission rates.}
    \end{figure}

Besides the production of elements, the nuclear heating from each reaction channel as a function of time is also highly dependent on the nuclear physics inputs.
In Figs.~\ref{fig:heatyesk} and~\ref{fig:fractionss1624} respectively, we show the total effective heating rates and the fractional effective heating rates for distinct theoretical nuclear models for simulations with $Y_e=0.24$ and $Y_e=0.16$. 
We remind the reader that we assume $M_{ej} = 0.05 M_\odot$ and $v_{ej}= 0.15c$ when estimating the thermalization efficiencies and in determining the total effective heating rate.
Similar to what was seen with the spread in predicted abundance patterns, Fig.~\ref{fig:heatyesk} shows that simulations with higher electron fraction show a relatively small spread in effective heating rates when compared to simulations with lower electron fractions. The initial electron fraction determines not only to what mass number the synthesis of heavy elements reaches, but also the places where there is a pile-up of nuclear species, and this has a direct impact on the extent to which different reaction and decay channels contribute to the overall heating.

As demonstrated in Fig.~\ref{fig:fractionss1624}, $\beta$-decay dominates the heating in most simulations leading up to roughly one day, which is a key time for KN energy generation.
From one to several days, these same simulations show significant competition between the contribution of spontaneous fission and $\alpha$-decay heating to the total heating. 
At 100 days, there is considerable variation in the effective heating rates for the lower $Y_e$ scenarios (Fig.~\ref{fig:heatyesk}(b)), despite the tendency for spontaneous fission heating to dominate at this time, as shown in Fig.~\ref{fig:fractionss1624}(b). 
We note that examining the fractional effective heating solely reflects relative contributions from decay processes and emphasize that the absolute heating from a given process like spontaneous fission can differ greatly between nuclear models since different fission barriers lead to different abundances of long-lived species such as \Cf{} \citep{Vassh2019Using}. 
The TF model shown in Fig.~\ref{fig:fractionss1624}(b) demonstrates such a case since here \Cf{} production is decreased relative to other models due to low TF fission barriers. 
We explore barrier effects on the populations of long-lived fissioning species further in Section~\ref{fission}. 
Lastly we note that although we mostly see late-time heating as being dominated by spontaneous fission and $\alpha$-decay, this is the case when we consider $\beta$-decay rates calculated based on \cite{Moller2019Nuclea}, as described in Section \ref{method:nucleosyn}.
When instead D3C$^*$ $\beta$-decay rates are adopted, we see markedly lower actinide production, thereby minimizing the heating impact from other decay channels and keeping $\beta$-decay heating as the dominant source throughout the calculation, yielding similar heating patterns as our higher $Y_{e}$ simulations, as shown by the dark purple line in Fig.~\ref{fig:heatyesk}(b).
Thus we caution that the full impact of processes such as spontaneous fission and $\alpha$-decay would have to be more carefully evaluated for a broader set of $\beta$-decay reaction rate calculations.

\begin{figure}[h!]
  \gridline{
  \fig{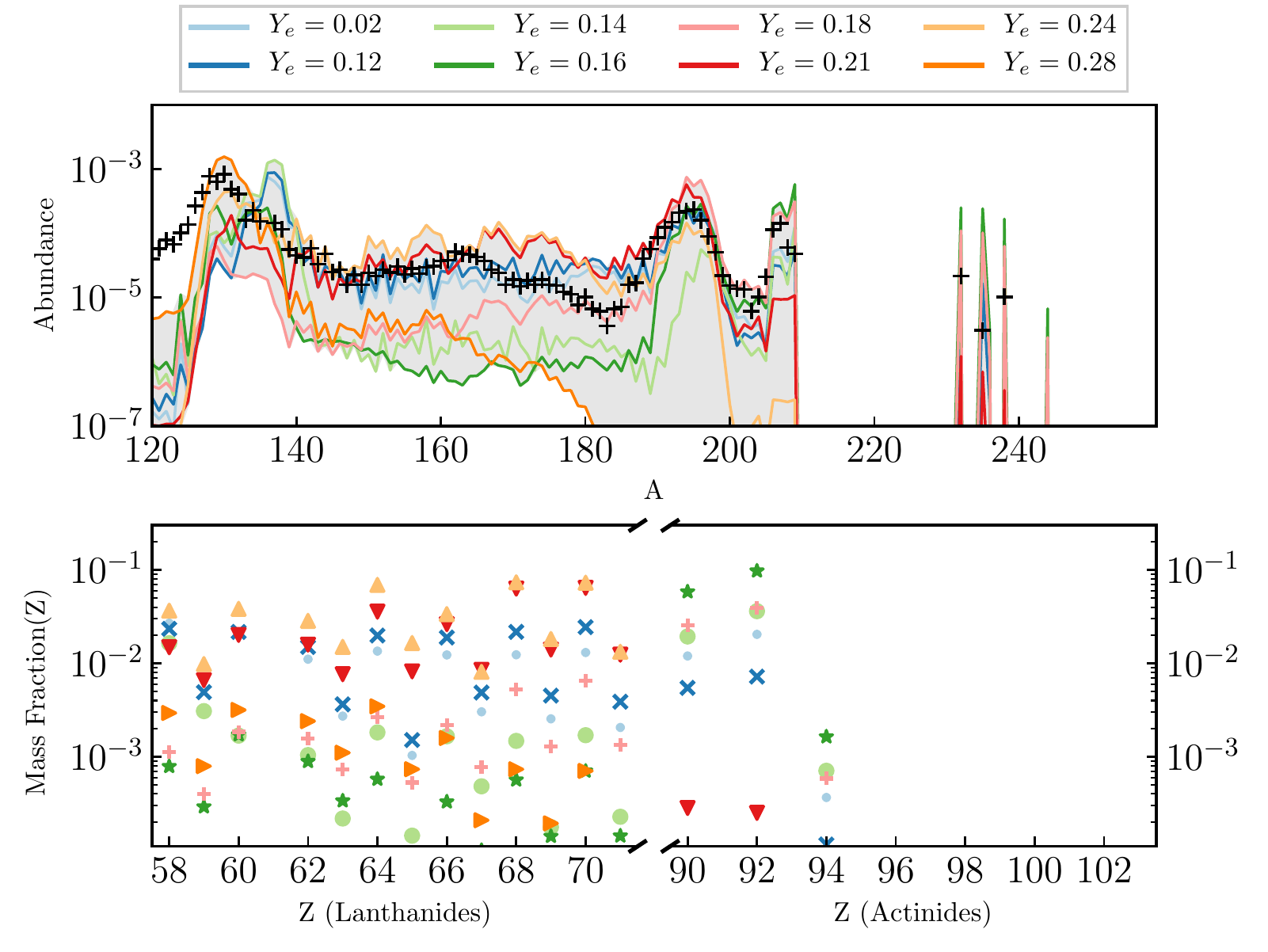}{0.49\textwidth}{(a) Simulations with HFB27 theoretical nuclear model.}             
  \fig{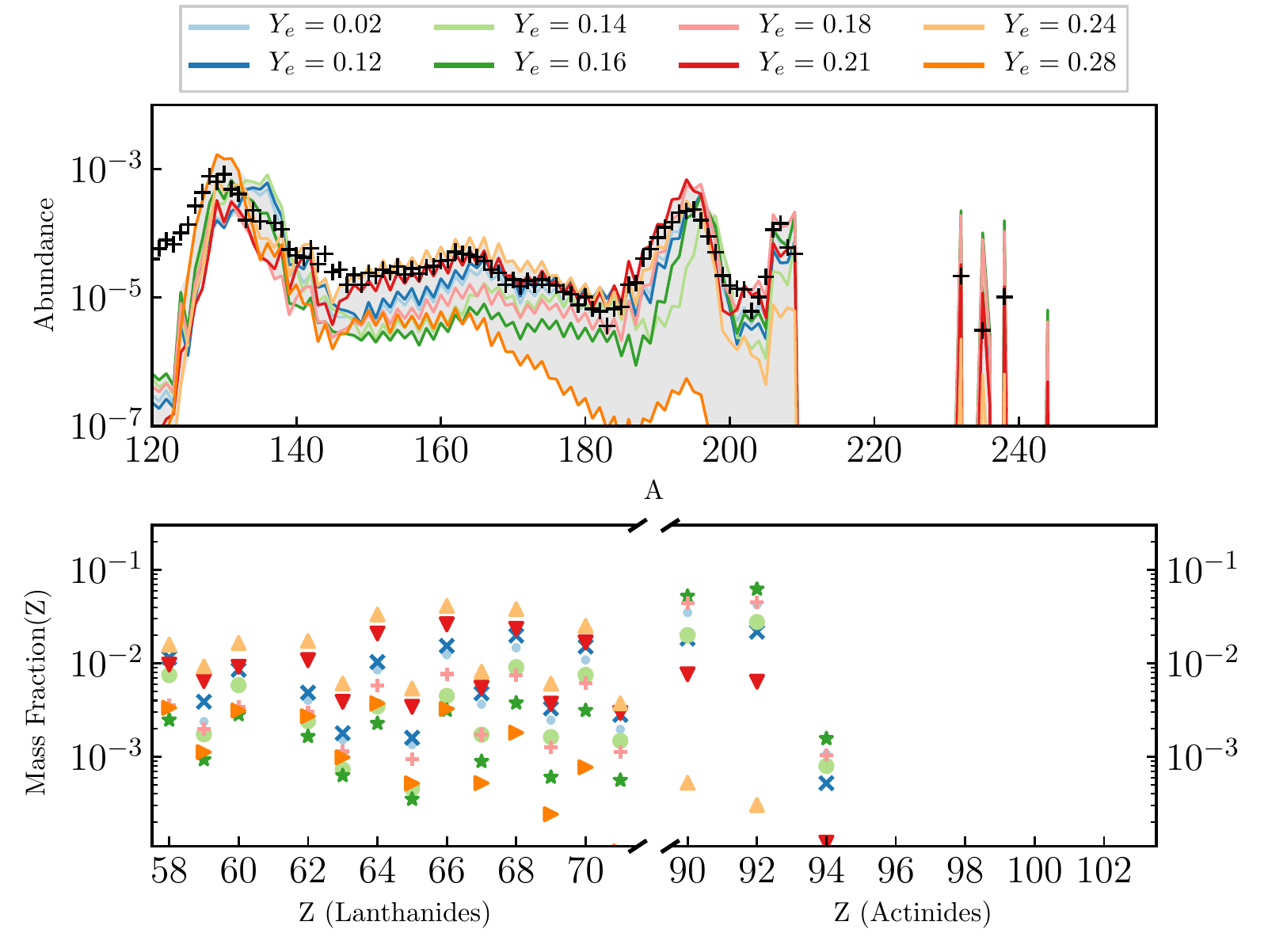}{0.49\textwidth}{(b) Simulations with FRDM2012 theoretical nuclear model.}}
  \gridline{
  \fig{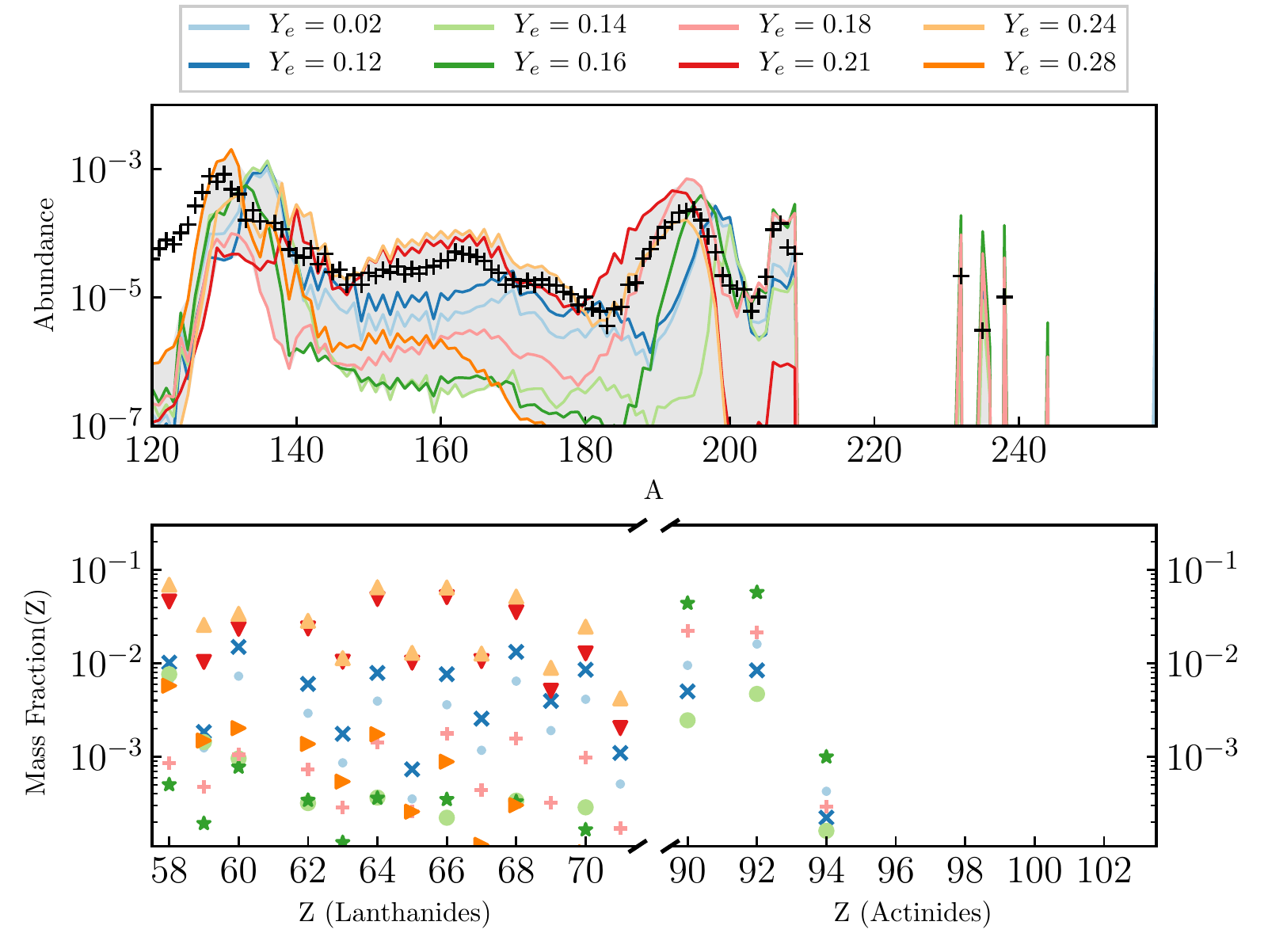}{0.49\textwidth}{(c) Simulations with UNEDF1 theoretical nuclear model.}
  \fig{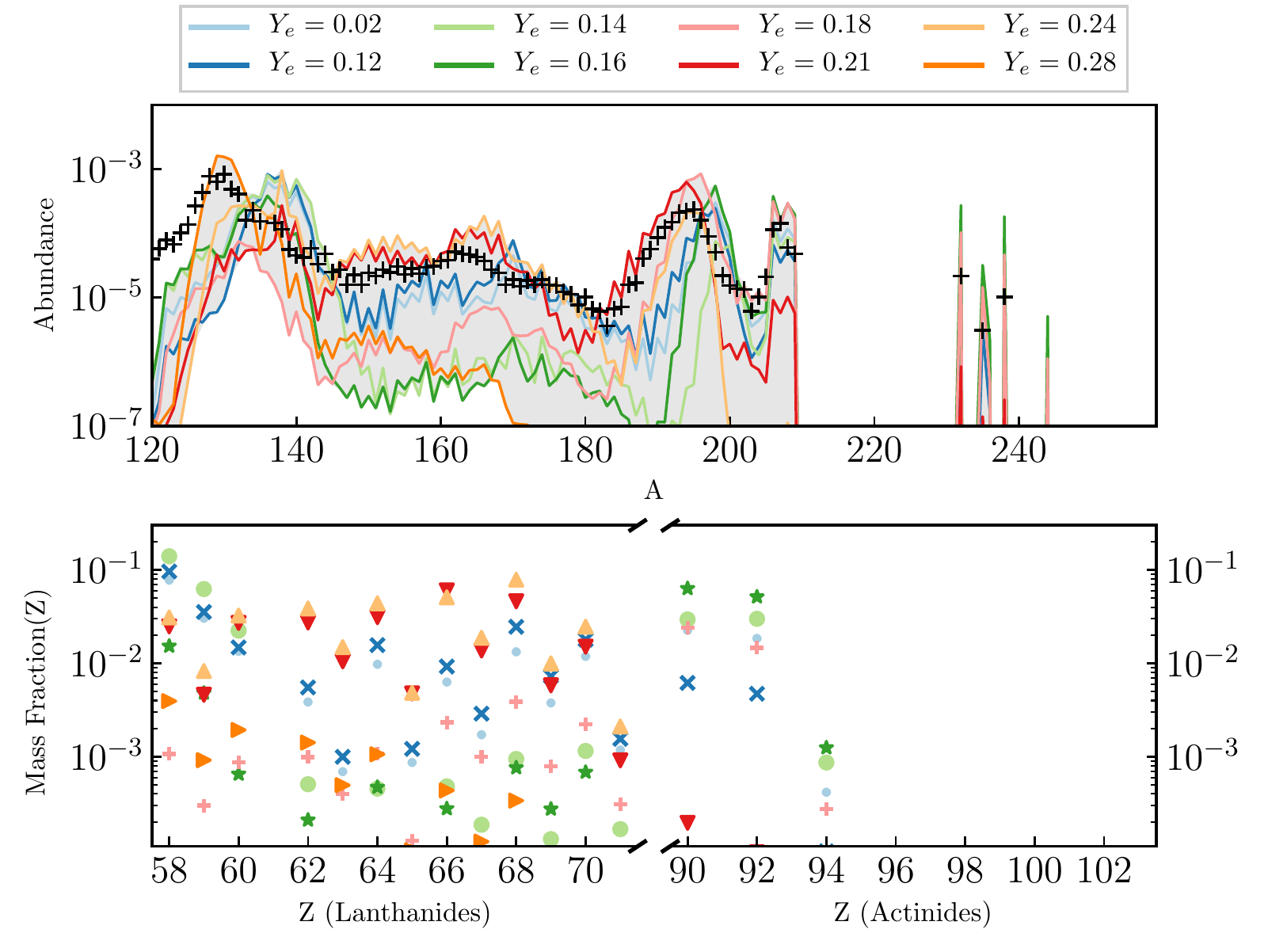}{0.49\textwidth}{(d) Simulations with ETFSI theoretical nuclear model.}}
  \caption{\label{fig:abmass}
  Final abundance pattern (top panels) and mass fraction of lanthanides and actinides (bottom panels) are shown for the same simulations as in Fig.~\ref{fig:heatyesk}.
  Black plus markers represent the solar \rp{} abundance.
  }
  \end{figure}
  \begin{figure}[h!]
  \gridline{
  \fig{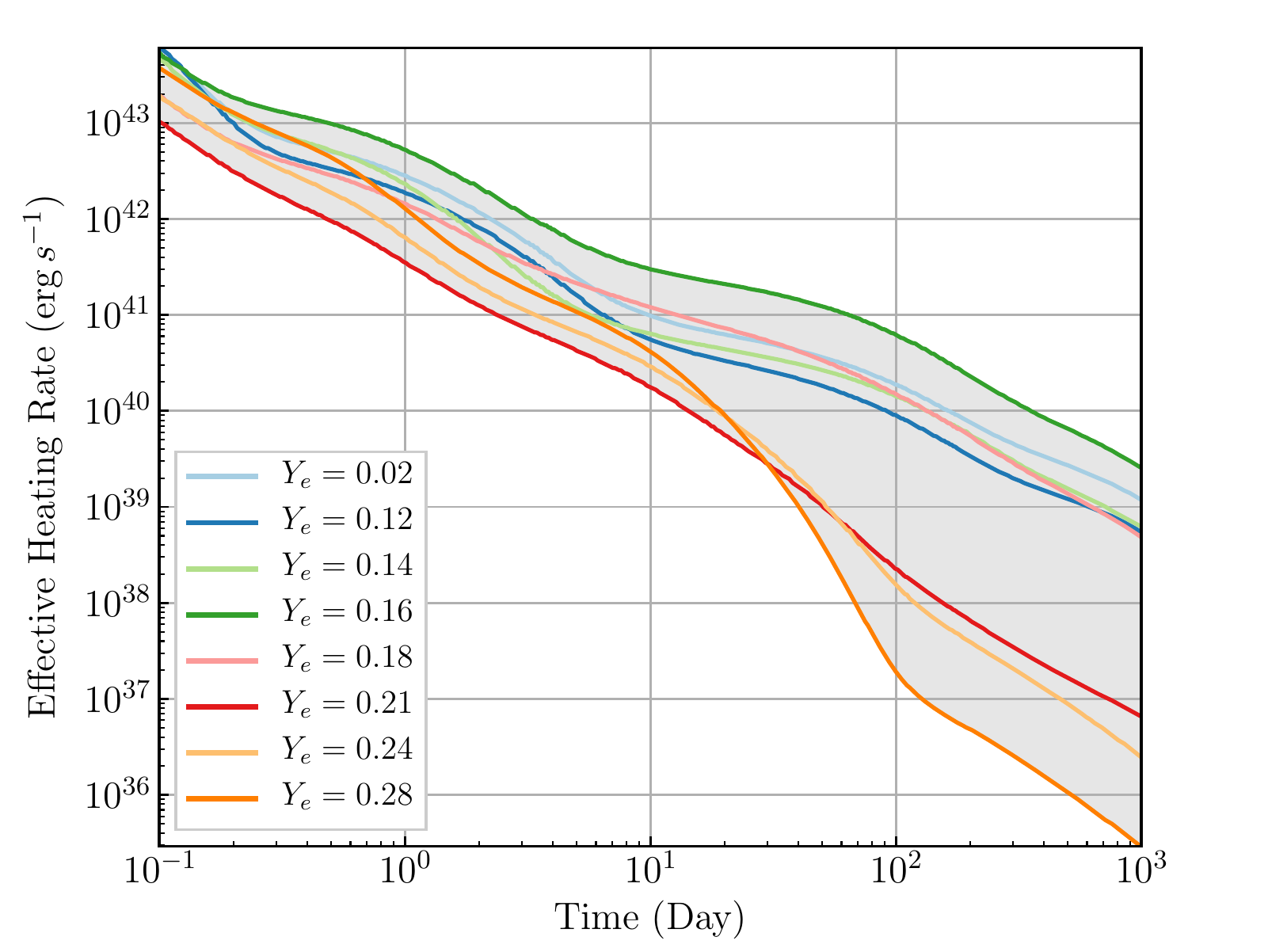}{0.49\textwidth}{(a) Simulations with HFB27 theoretical nuclear model.}              
  \fig{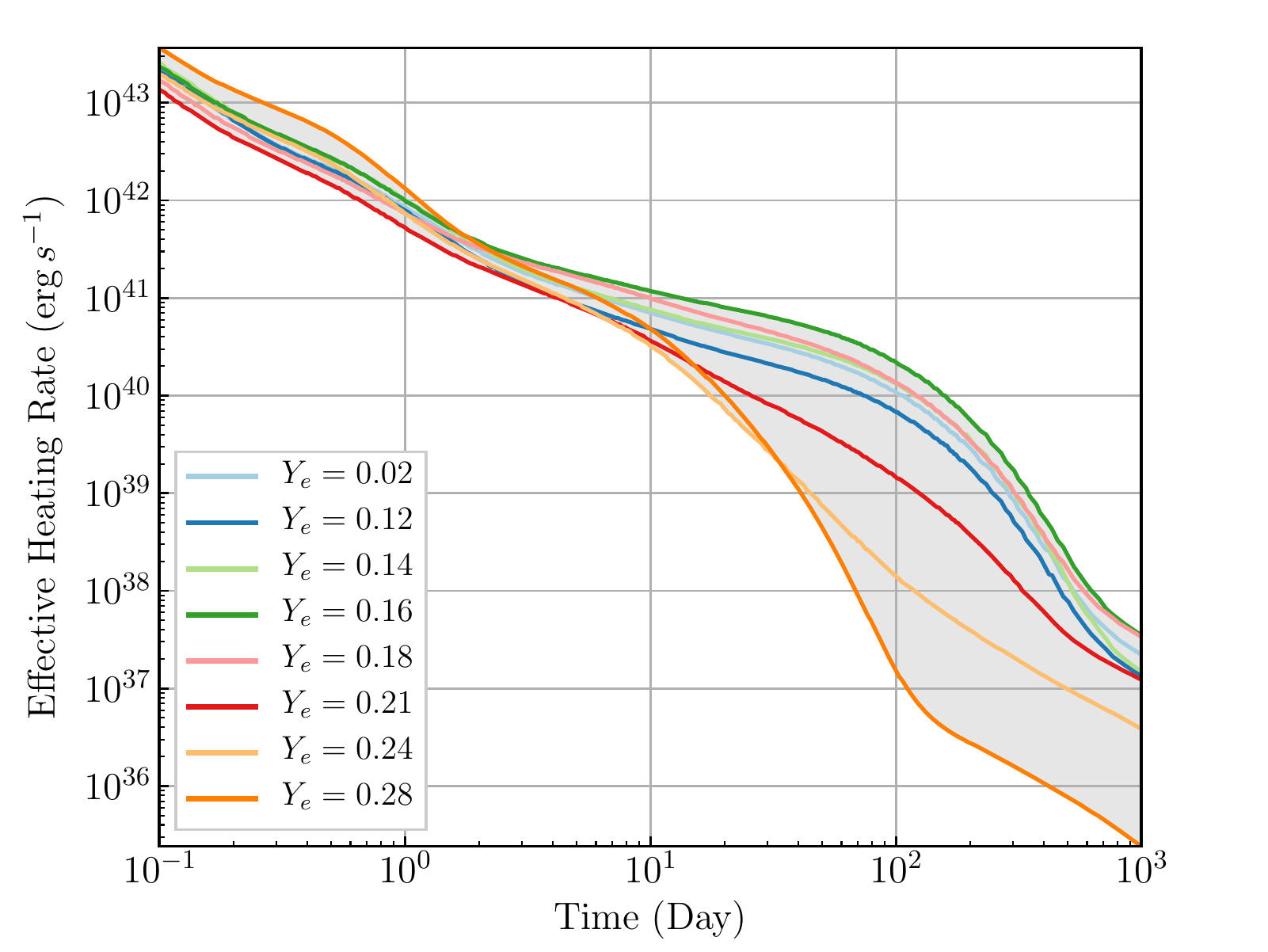}{0.49\textwidth}{(b) Simulations with FRDM2012 theoretical nuclear model.}}
  \gridline{
  \fig{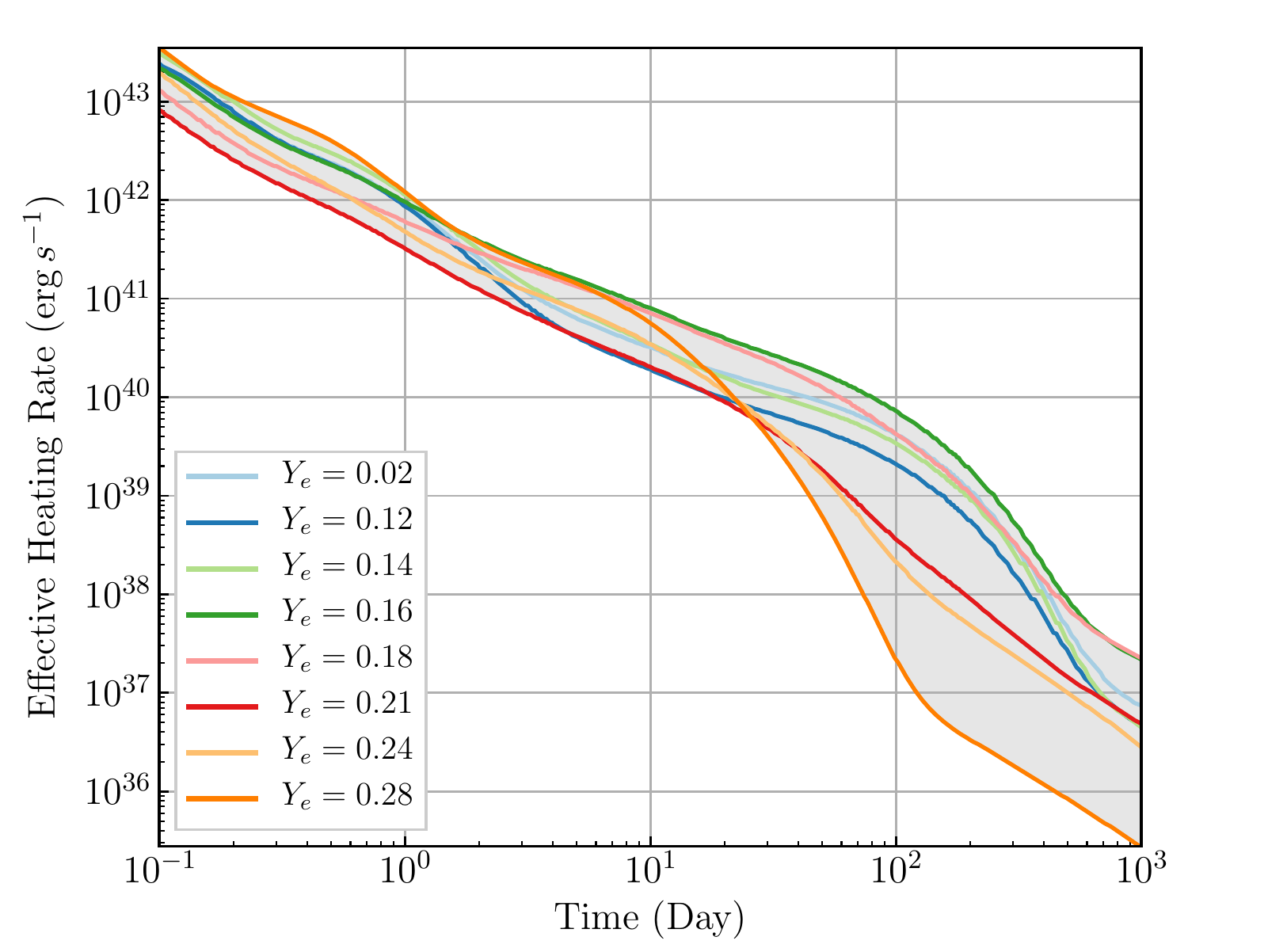}{0.49\textwidth}{(c) Simulations with UNEDF1 theoretical nuclear model.}
  \fig{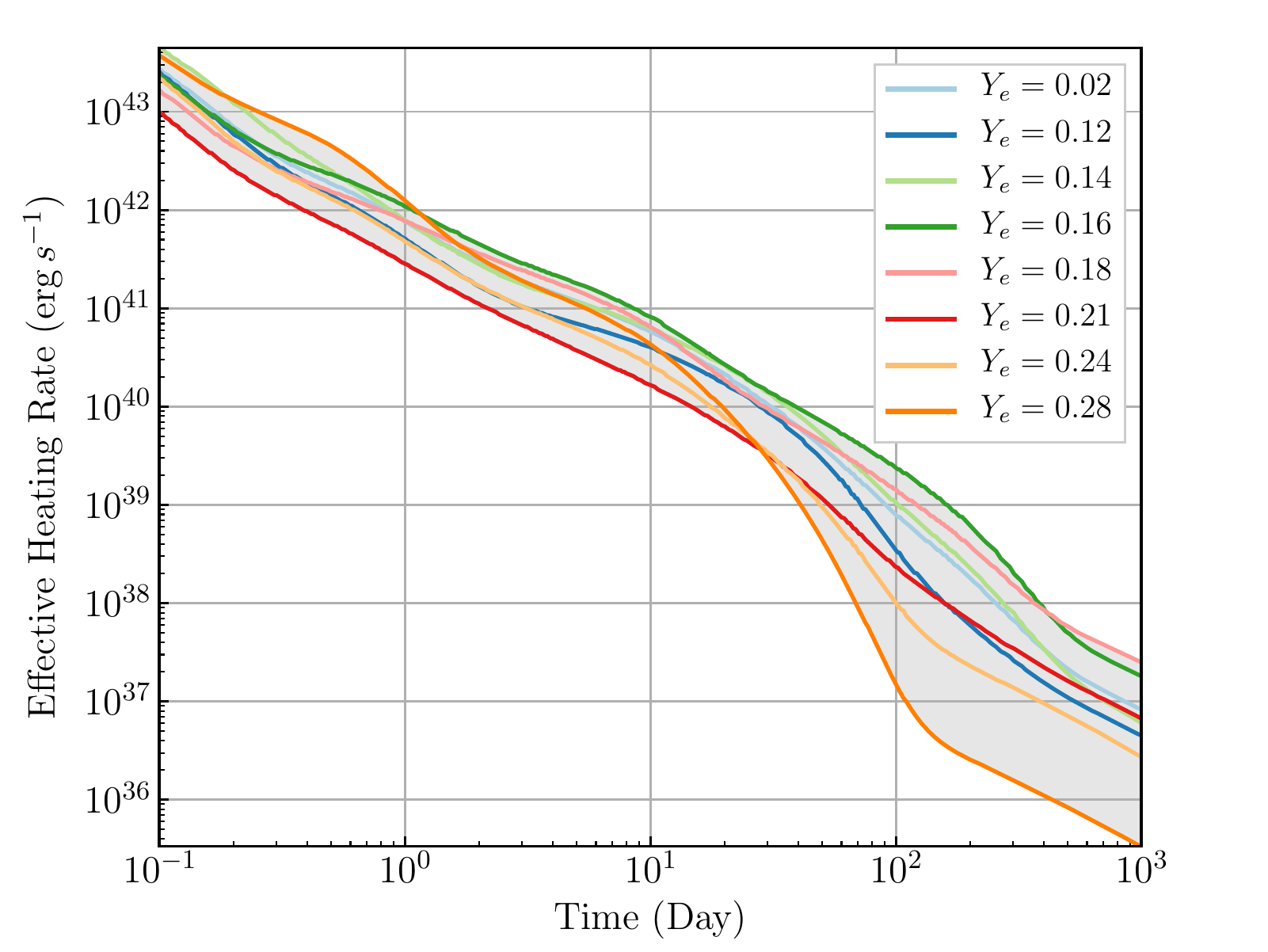}{0.49\textwidth}{(d) Simulations with ETFSI theoretical nuclear model.}}
  \caption{\label{fig:heatmasssk}
  Total effective heating rates for various electron fractions ($Y_e$).
  These simulations use symmetric fission yields and Karpov (KZ) spontaneous fission rates.
  The grey band shows the range of effective heating rates from all simulations.
  }
  \end{figure}

We now turn to an illustration of the effect of varying the astrophysical conditions for a fixed theoretical nuclear mass model. 
Fig.~\ref{fig:abmass} shows the comparison of four nuclear models, and for each we use a set of simulations with electron fractions ranging from $Y_{e}$ of 0.02 to 0.28 as described in Section~\ref{method:astro}.
We compare the final abundance pattern as well as lanthanide and actinide mass fractions for the four sets of simulations.
We note that each simulation displayed in Fig.~\ref{fig:abmass} shows production of \rp{} nuclei beyond the second \rp{} peak (at $A\sim130$).
The grey bands in each subplot show the range of the final abundance patterns for each set of simulations.  
The size of the spread of predictions varies with mass model, with FRDM2012 (and FRLDM fission barriers) having only about a one order of magnitude range of final abundance in the lanthanide region. 
However, the same region has a two order of magnitude spread of predictions with each of the theoretical nuclear models  HFB27, UNEDF1, and ETFSI.

The third \rp{} peak~(at $A\sim195$) and beyond shows the largest sensitivity to the initial electron fraction. 
Consistent with Fig.~\ref{fig:heatyesk}(a), we see in Fig.~\ref{fig:abmass} that the abundance of nuclei at and beyond the third peak is very low in the simulations with an initial electron fraction above roughly ${\rm Y_e} \sim  0.24$.
Meanwhile, we tend to see maximal production beyond the third \rp{} peak and correspondingly low final abundance in the rare-earth region in simulations that have just enough neutrons to capture out significantly past the third peak i.e. initial electron fractions of 0.16 to 0.18.  
The final abundance pattern in the lanthanide region is again larger for even lower electron fractions ($Y_e < 0.14$) since in these cases, significant fission deposition is taking place as well as subsequent neutron capture on the fission products.

In Fig.~\ref{fig:heatmasssk} we show the spread of effective heating rates corresponding to the simulations in Fig.~\ref{fig:abmass}. 
As with the final abundance pattern, we see the least variation in the heating in the simulations using FRDM2012 nuclear models. 
For the remaining models we see that there is not a one-to-one correspondence between the spread in the heating rate and the spread in the predicted abundance pattern.
This is because the effective heating is dependent upon the energy released from each reaction and decay channel coupled with the relevant thermalization efficiencies, and the amount of heating in each channel can vary with nuclear inputs.

\section{Effects of Uncertainties in KN Light-Curve Modeling}\label{uncertainties}
\begin{figure}[!]
  \gridline{
  \fig{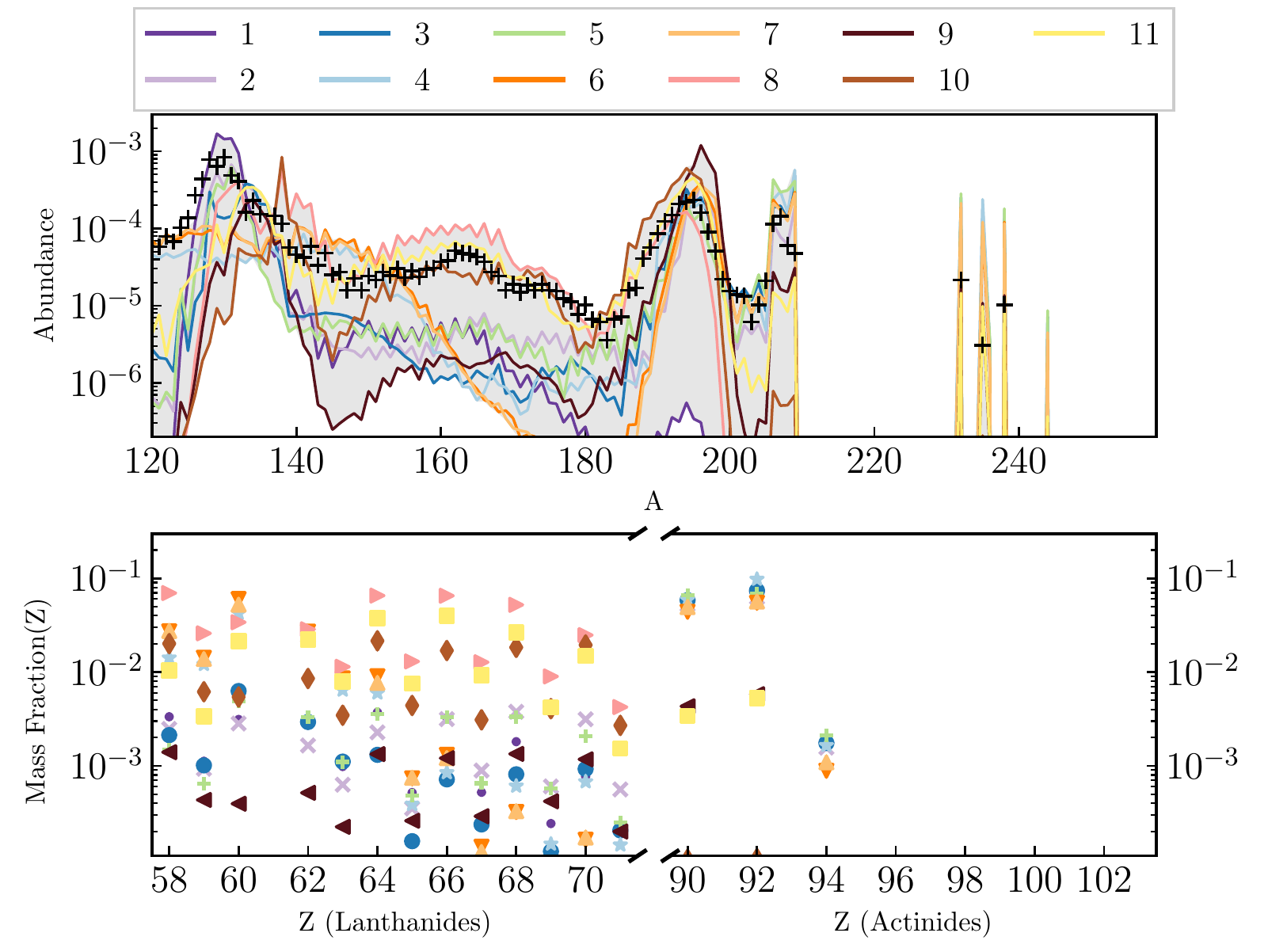}{0.49\textwidth}{(a) 11  single simulations, labeled $1$ to 11 in Table.~\ref{tab:13calc}.}
  \fig{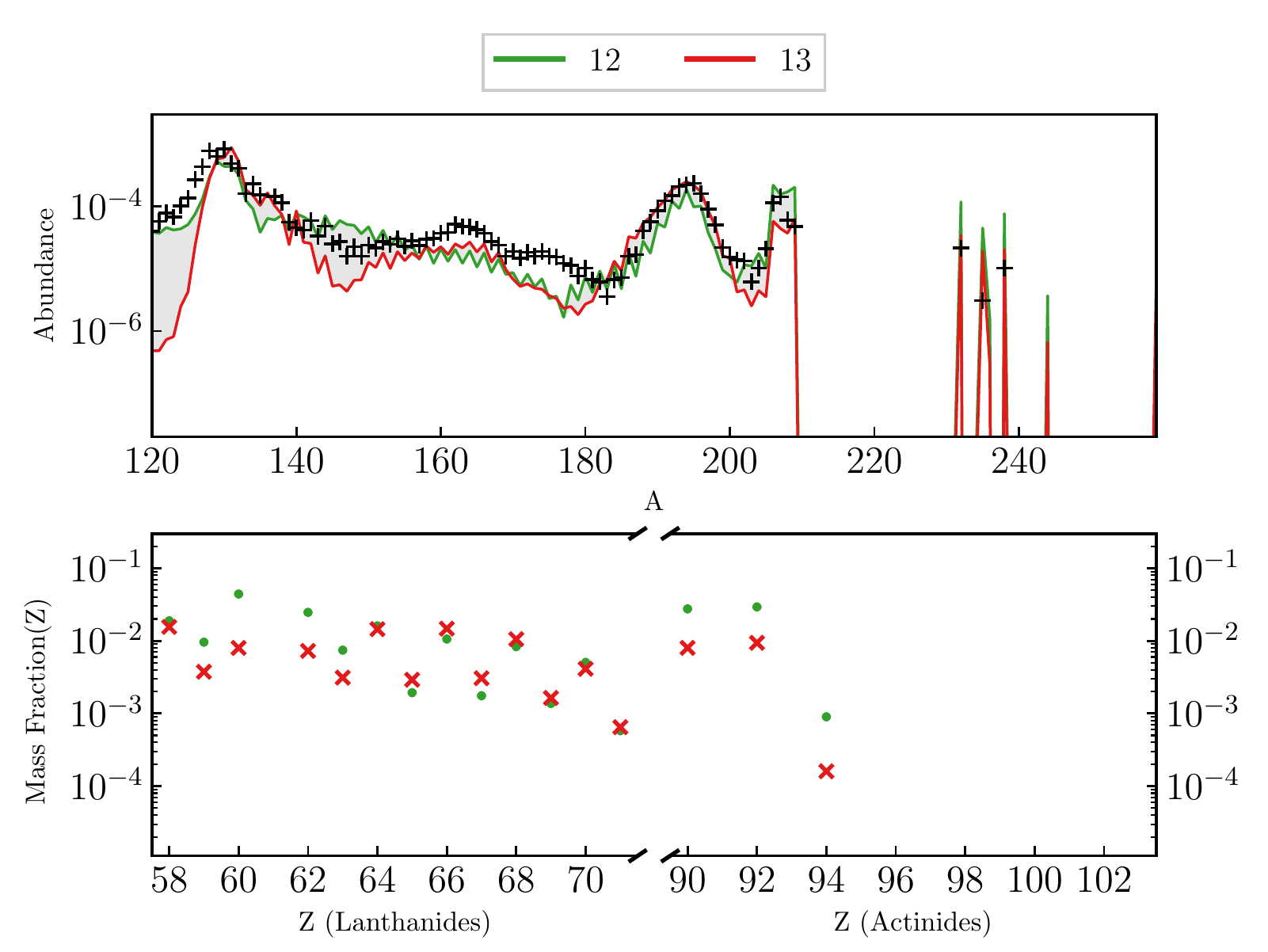}{0.49\textwidth}{(b) Two cases with a linear combination of simulations to reproduce the solar abundance pattern where $120 <  A < 200$, labeled as 12 and 13 in Table.~\ref{tab:13calc}.}}
  \caption{\label{fig:aba}The final abundance (top panel), mass fractions of individual lanthanide and actinide elements (bottom panel) from the nucleosynthesis simulations. 
  For comparison, the solar abundance is plotted with black plus markers.
  The grey band (top panel) and grey vertical lines (bottom panel) show the range of final abundance and mass fractions of individual lanthanide and actinide elements from all simulations included in each figure.
  }
\end{figure}

In this section, we explore the effect of the range of uncertainties in composition and effective heating rates described in previous sections on the KN light curve. 
In order to sample from the full range of predicted heating rates and final abundances from all our simulations, we select 13 simulations, which are listed in Table~\ref{tab:13calc}.  In choosing the simulations labeled 1-11 in Table~\ref{tab:13calc}, we include reasonable variation in theoretical nuclear model, spontaneous fission prescription, and initial $Y_{e}$. 
A subset of these models have the same $Y_{e}$, which allows us to better explore the effect of changing nuclear physics inputs.
The simulations labeled 12  and 13  are a linear combination of simulations with different initial electron fractions designed to match the solar abundance pattern (indicated by black crosses in Fig.~\ref{fig:aba}). 
As shown in Fig.~\ref{fig:aba}(b), simulation 12 fits the solar pattern in the second and third peaks, with an overproduction around $A = 150$, while simulation 13 fits the solar pattern well from the second peak out past the third peak.
We find that together, all thirteen selected simulations adequately capture the complete span of effective heating curves, as can be seen in Fig.~\ref{fig:effectiveheatband}.

\begin{deluxetable}{c ccccc}
  \tabletypesize{\footnotesize}
  \tablewidth{0pt}
  \tablecaption{\label{tab:13calc}Highlighted simulations with their mass models, initial electron fractions and fission prescriptions}
  \tablehead{
  \colhead{Index} & \colhead{Mass Model} & \colhead{$Y_e$ } & \colhead{Yield Distribution} & \colhead{Spontaneous Fission rates} & \colhead{$X_{\rm Ln+An}(t_\kappa = 1 \; {\rm day})$}
  }
  \startdata
  1 & FRDM2012 & 0.28 & Symmetric & KZ & $0.020$\\
  2 & FRDM2012 & 0.16 & Symmetric &  KZ & $0.217$ \\  
  3 & HFB22 & 0.16 &  Symmetric & KZ & $0.366$ \\ 
  4 & HFB27 & 0.16 &  Kodama & KZ & $0.481$\\
  5 & DZ33 & 0.16 & Symmetric & KZ & $0.308$ \\
  6 & UNEDF1 & 0.16 & Kodama & KZ & $0.336$\\
  7 & UNEDF1 & 0.16 & Kodama & XR & $0.335$\\
  8 & UNEDF1 & 0.24 & Symmetric & KZ & $0.378$\\
  9 & SLY4 & 0.18 & Symmetric & KZ & $0.029$\\  
  10 & SLY4 & 0.21 & Symmetric & KZ & $0.120$\\  
  11 & TF$+$D3C$^\ast$ & 0.16 & Symmetric & KZ & $0.217$ \\  
  12 & DZ33 & mixture\tablenotemark{a}  & Kodama & KZ & $0.258$\\
  13 & UNEDF1 & mixture\tablenotemark{b} & Symmetric & KZ &$0.115$
  \enddata
\tablenotetext{a}{Linear combinations of DZ33 simulations with $Y_e$ values of $0.02$,$0.14$,$0.16$, $0.18$, $0.28$ and the corresponding weights of $38\%$, $6.8\%$, $20\%$, $8.6\%$, $26\%$.}
\tablenotetext{b}{Linear combinations of UNEDF1 simulations with $Y_e$ values of $0.02$, $0.16$, $0.18$, $0.21$, $0.24$, $0.28$ and weights of $8.2\%$, $7.2\%$, $18\%$, $25\%$, $1.4\%$, $40\%$.}
\end{deluxetable}

\subsection{Nucleosynthesis and Effective Heating}\label{sec:nucheat}
\begin{figure}[h!]
  \centering
   \includegraphics[width = 0.7\textwidth]{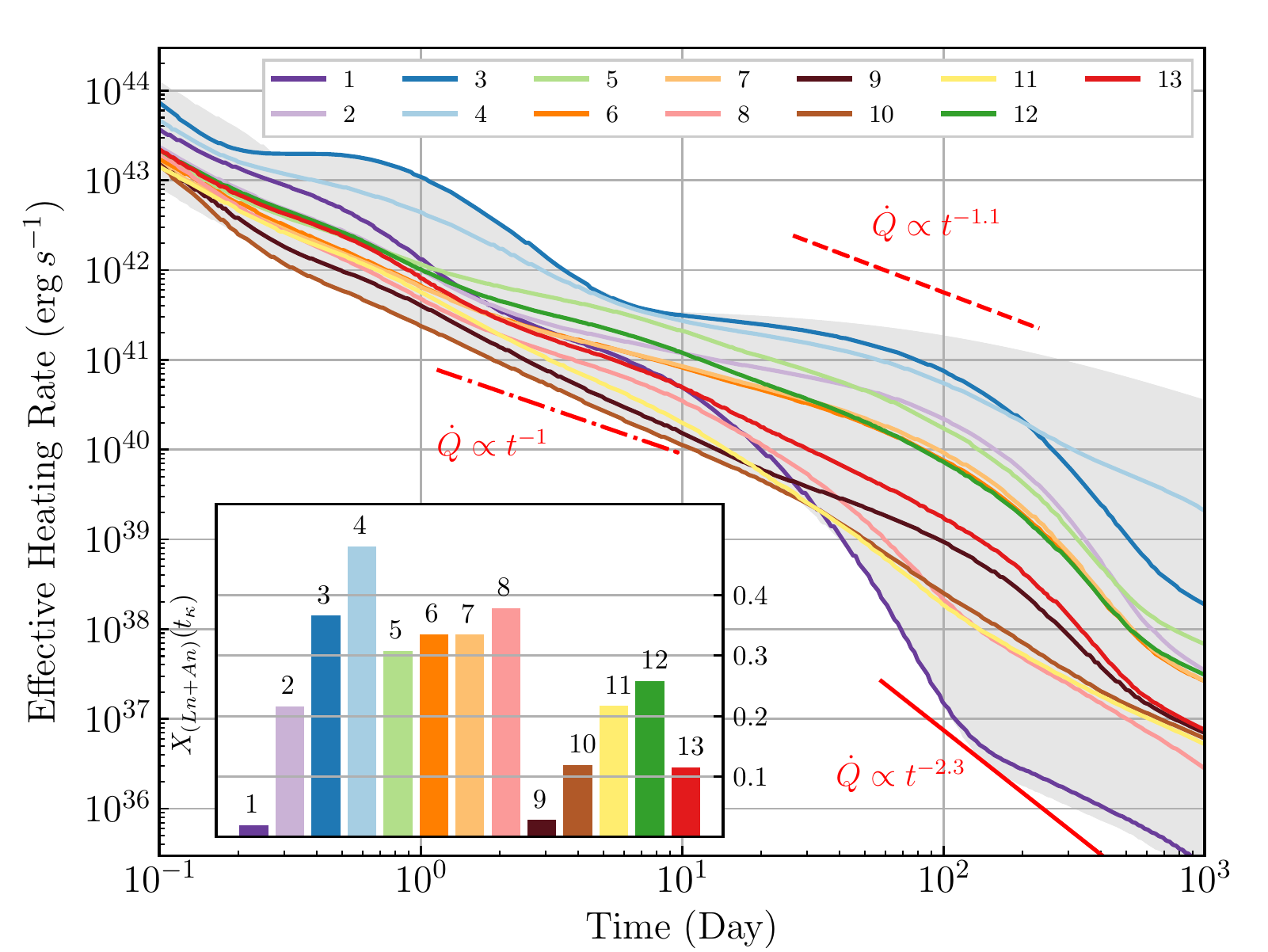}
\caption{\label{fig:effectiveheatband}Comparison of the effective heating rates of the 13 simulations in Table~\ref{tab:13calc}. Subplots in the left bottom are mass fractions of lanthanides ($Ln$) and actinides ($An$), $X_{\rm Ln+An}(t_\kappa)$, where $t_\kappa=1$ day. Lines proportional to $t^{-\alpha}$ are shown for comparison.
The grey band shows the range of effective heating rates from all our astrophysical and nuclear inputs.}
\end{figure}

\begin{figure}[!ht]
\centering
\includegraphics[angle=0,width = 0.59\textwidth]{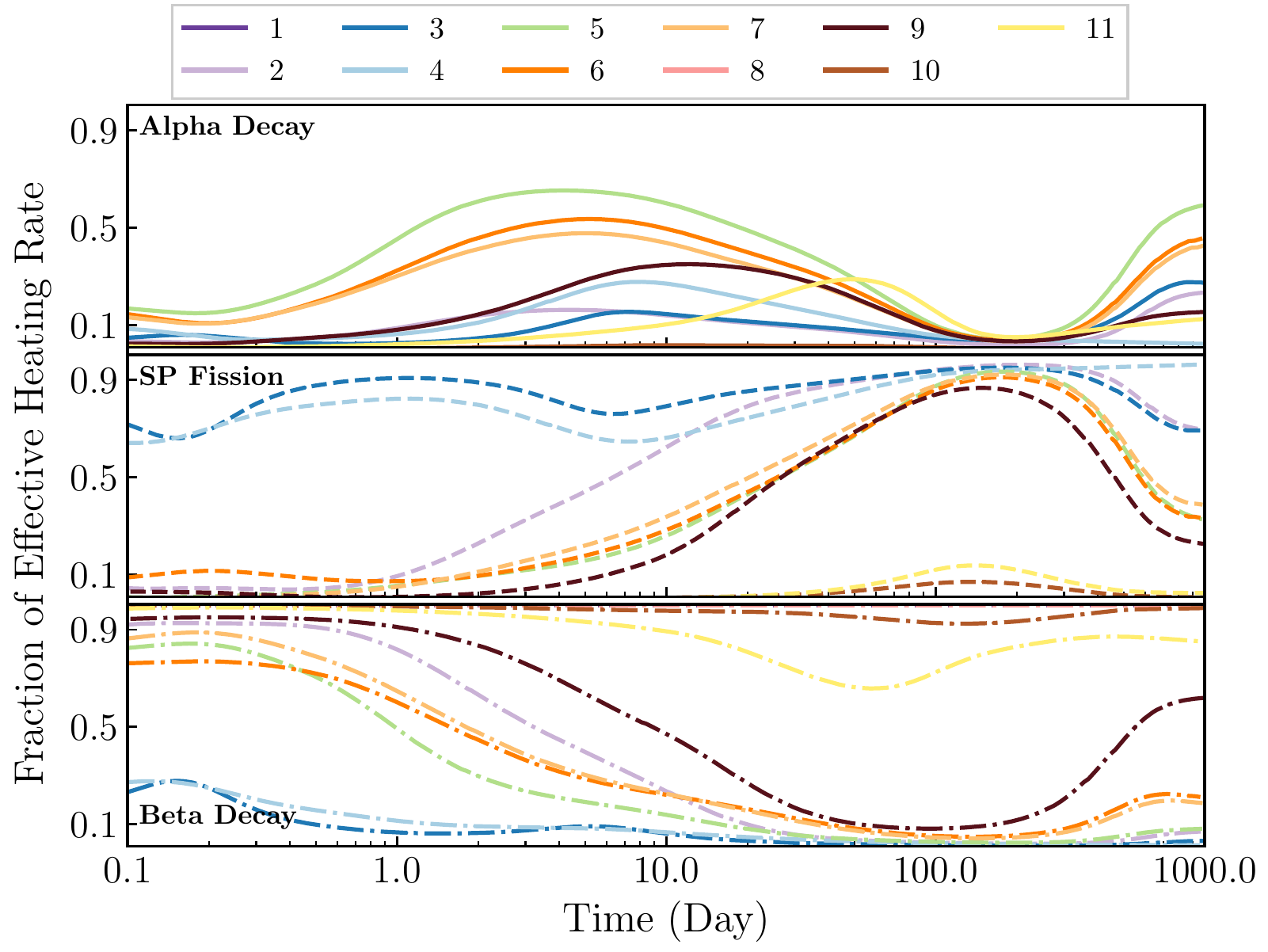}
\caption{\label{fig:10fraction}Fractional heating rates for different nuclear channels using individual simulations, labeled in Table~\ref{tab:13calc}.}
\end{figure}

\begin{figure}[!h]
\gridline{
\fig{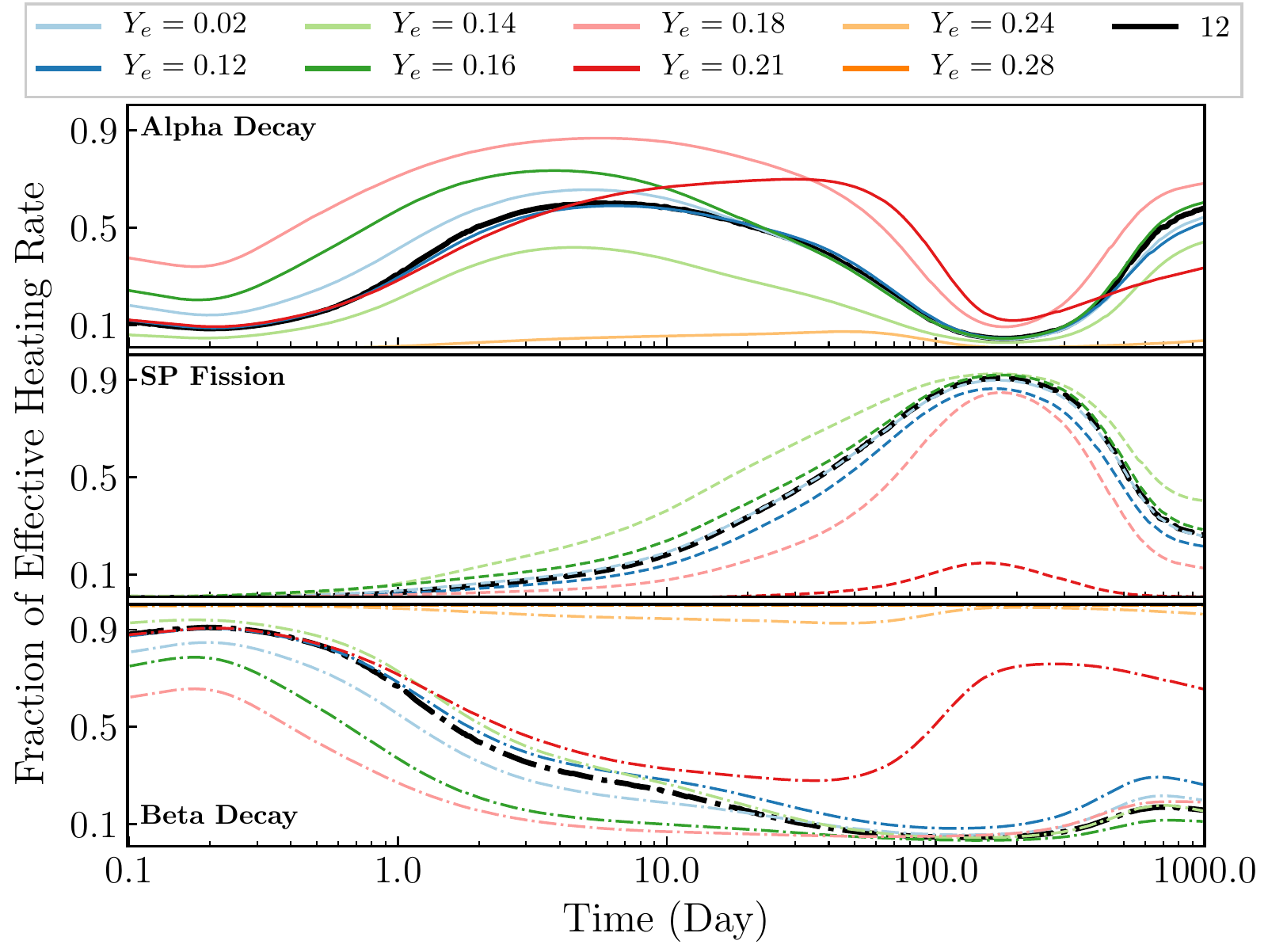}{0.49\textwidth}{(a) Simulation 12 which uses the DZ33 theoretical nuclear model and Kodama fission yields.} 
\fig{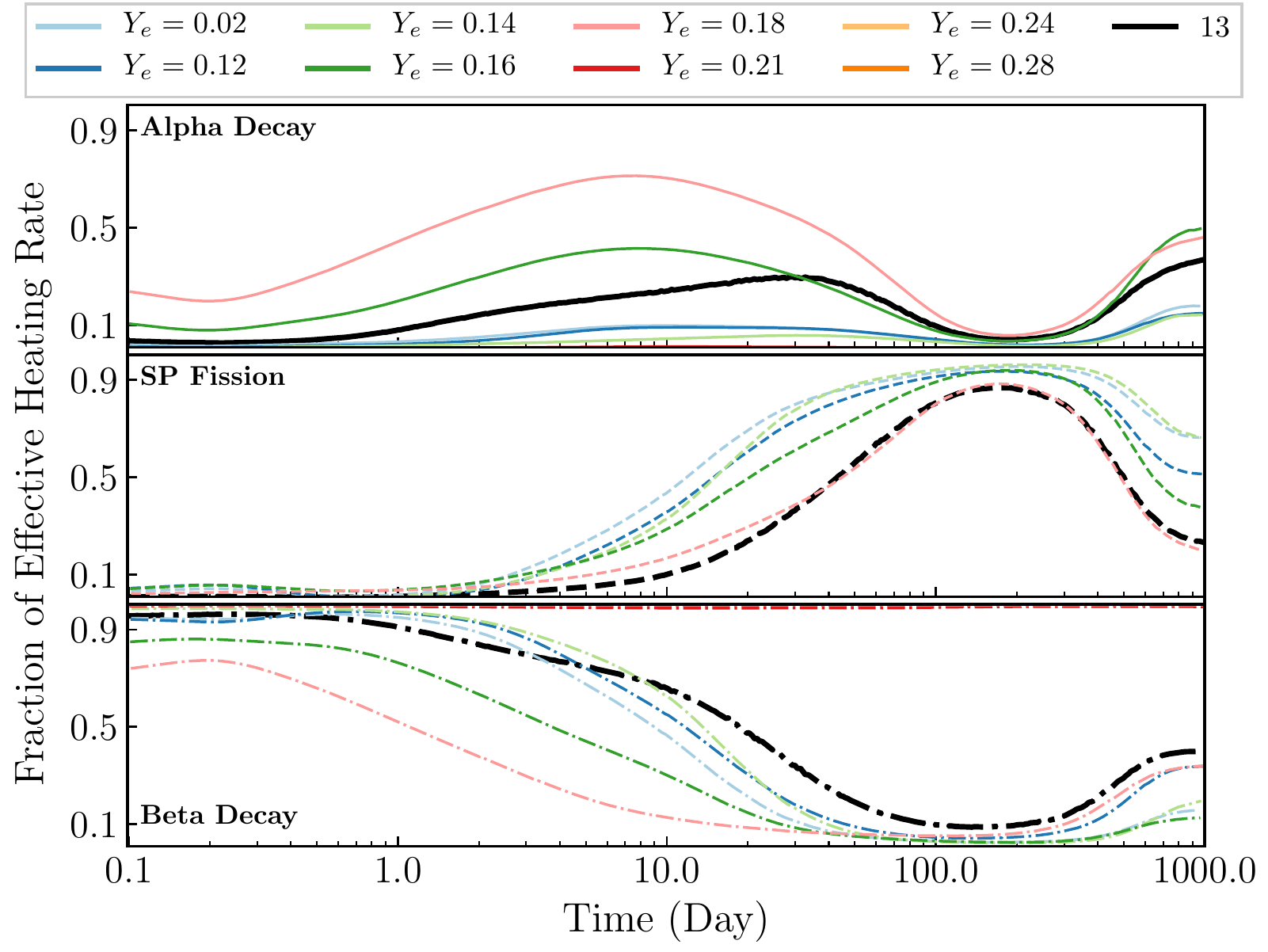}{0.49\textwidth}{(b) Simulation 13 which uses the UNEDF1 theoretical nuclear model and symmetric fission yields.}}
\caption{\label{fig:fracmix}Fractional heating rates of the two linear combinations of simulations that both reproduce the solar abundance pattern in the range of $120 < A < 200$.
Colored lines represent the individual simulation from which the linear combination is composed.
Black lines represent simulation 12 (left) or 13 (right).
Both simulations use the Karpov (KZ) spontaneous fission rates.}
\end{figure}

In Fig~\ref{fig:effectiveheatband}, we show the overall effective heating rate for the 13 simulations, calculated using the thermalization efficiencies described in Section~\ref{method:thermalization} with $M_{ej} = 0.05 M_\odot$ and $v = 0.15c$. 
In this figure, the grey band represents the range of effective heating rates from all simulations with $Y_{e}$ of 0.02 to 0.28, as well as the full set of nuclear physics inputs described in Section~\ref{method:nucleosyn}.
The combined mass fraction of lanthanides and actinides, $X_{\rm Ln+An}(t_\kappa = 1 \; {\rm day})$, varies from 0.020 to 0.481, as shown in the corner subplot of Fig.~\ref{fig:effectiveheatband}.
In Fig~\ref{fig:10fraction}, we show the fractional effective heating rates for the individual simulations 1-11.
Meanwhile, Fig.~\ref{fig:fracmix} shows the fractional effective heating rates for simulations 12 and 13, in addition to the individual simulations of which they are composed.


Prior to one day post-merger, we see that most simulations are dominated by $\beta$-decay. 
The exceptions to this are simulations 3 and 4, which use the HFB theoretical nuclear model. 
In these cases, Fig.~\ref{fig:10fraction} shows the dominance of spontaneous fission at this time, and we see a correspondingly high effective heating rate, shown in Fig.~\ref{fig:effectiveheatband}. 
Beginning at 8-10 days post-merger, we start to see $\alpha$-decay and spontaneous fission playing an increasingly important role in the overall heating for simulations with $Y_{e}<0.21$, as well as those fit to the solar pattern. 
We note that these cases are weighted such that they include substantial contribution from simulations with $Y_{e}<0.21$ in order to produce solar actinide yields.
Around this time, we see that the effective heating rates for individual simulations decay with slopes roughly proportional to $t^{-1}-t^{-1.1}$. 
For comparison purposes, these timescales are shown by the red lines in Fig.~\ref{fig:effectiveheatband}.
Simulations 1, 8 and 10 ($Y_{e}\geq0.21$) lack significant contributions from spontaneous fission and $\alpha$-decay, and show lower heating rates than their lower $Y_{e}$ counterparts. 
However we also see that for a given $Y_{e}$, the effective heating rate is also sensitive to the theoretical nuclear model.

Around 100 days, simulations 1, 8 and 10 continue to be dominated by $\beta$-decay. 
The effective heating rate for these simulations decays rapidly at this time, roughly proportional to $t^{-2.3}$, as indicated by the solid red line in Fig.~\ref{fig:effectiveheatband}.
Meanwhile, simulations with low $Y_{e}$, as well as simulations 12 and 13, are dominated by spontaneous fission and in most cases show a \lq\lq bump" in the overall heating compared to the power law decays.
This \lq\lq extra" effective heating mainly comes from the spontaneous fission of \Cf, which is consistent with the findings of~\citet{Zhu2018Califo}.
After hundreds of days, we tend to see a decreased importance of spontaneous fission and a corresponding rapid decay in the effective heating rate. 
We further explore the sources and roles of theses different processes and timescales in energy generation in Section~\ref{fission}. 

\subsection{Light Curves}\label{sec:curves}

\begin{figure}[h!]
  \centering
   \includegraphics[width = 0.75\textwidth]{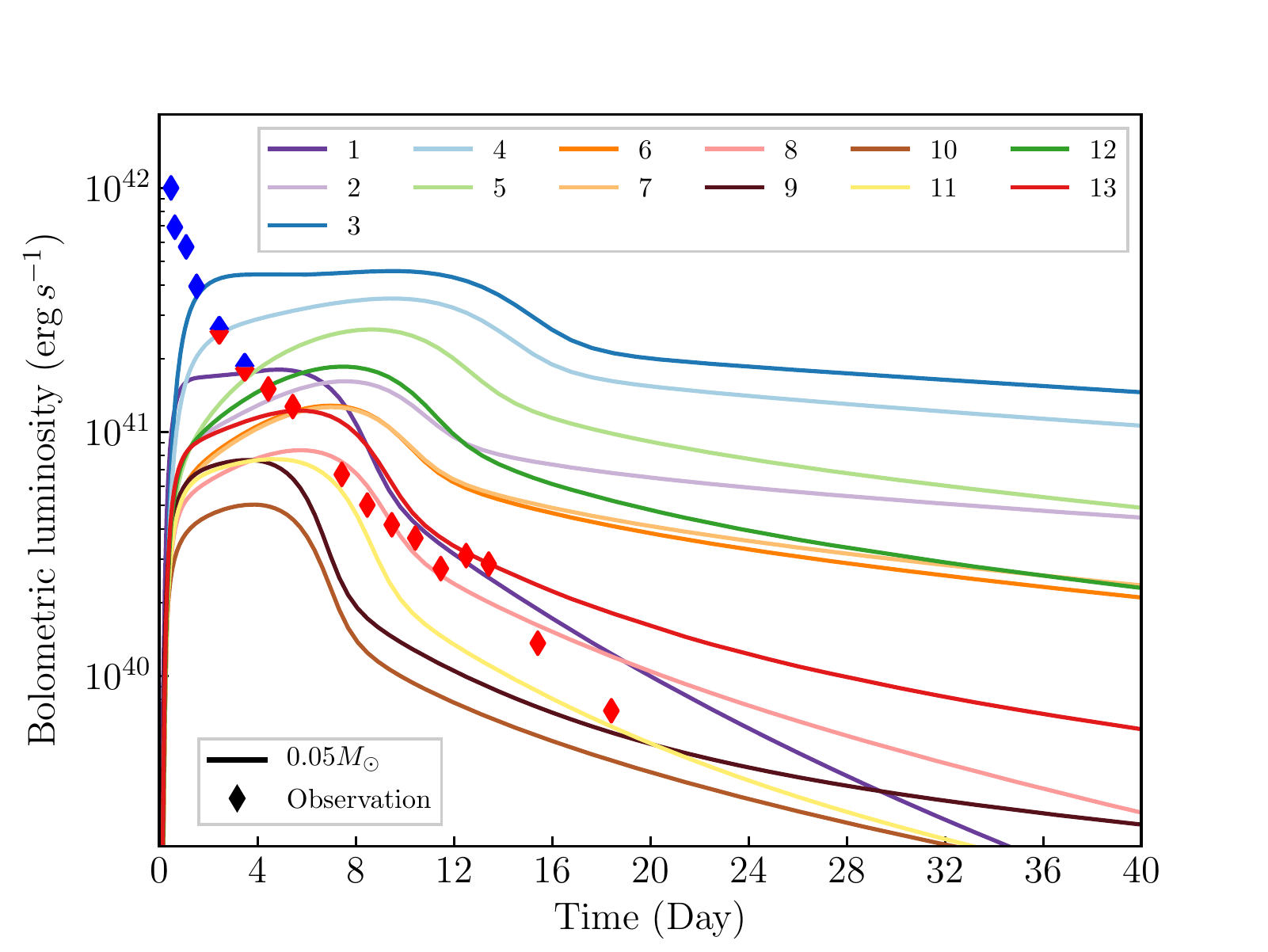}
\caption{\label{fig:bolol}
Bolometric luminosity of 13 simulations with the ejecta mass $0.05\,M_{\odot}$ and the ejecta velocity $v_{ej}=0.15c$. 
Red and blue diamond markers are observed bolometric luminosity from the KN associated with GW170817~\citep{Drout2017Light}.}
\end{figure}

The relationship between \rp{} radioactivity and the emerging light curve is governed by multiple interacting factors. 
While the total effective heating sets the overall magnitude of the light curve, the maximum luminosity also depends on the time at which the light curve peaks, which is a function of the ejecta opacity. 
Changes in the slope of the effective heating curve can also affect the shape of the light curve near its peak, while the decline of the light curve is a function of late-time effective heating, which is directly tied to nucleosynthetic outcomes, particularly the prevalence of alpha decay and fission.
We now take the thirteen models given in Table~\ref{tab:13calc}, and calculate the bolometric luminosity for each of them as described in Section~\ref{sec:method}.  
Fig.~\ref{fig:bolol} shows the results of this calculation assuming an ejecta mass of $M_{\rm ej} = 0.05 \, M_\odot$ and an ejecta velocity of $v_{\rm ej} = 0.15c$.
For comparison, the  observed bolometric luminosity from the KN associated with GW170817~\citep{Drout2017Light} is also shown.
This includes emission from a blue component, which dominates for the first ${\sim}2$ days, and a red component, which is the main contributor after roughly 4 days. 
Since our model is concerned only with the red component, the relevant time period for comparison with data is $t \gtrsim 4$ days.
As can be seen in Fig. \ref{fig:bolol}, the peak bolometric luminosity is on the scale of $10^{41}$~erg s$^{-1}$, and all simulations reach their maximum values by times ranging from 2 days to about 12 days. 
We note that this behavior is for an ejecta mass close to that generally inferred from GW170817. We explore the effects of varying the ejecta mass in the next section.  

Much of the variability in the luminosity is a reflection of the variability in the effective nuclear heating seen in Fig.~\ref{fig:effectiveheatband}. 
For the selected simulations, the range of uncertainty in the peak bolometric luminosity is from $5\times 10^{40}$ to $3\times 10^{41}$ erg s$^{-1}$. 
The general trend is that a larger effective heating rate at the peak time leads to an overall more luminous light curve.  
For example, the effective heating rates of simulation 3 and 4 are close to the upper bound of the effective heating rate range shown in  Figure~\ref{fig:effectiveheatband} and the peak bolometric luminosities of simulations 3 and 4 shown in Figure~\ref{fig:bolol}  are also the most luminous.  
At a few days after the merger, simulations 9 and 10 have the lowest effective heating rate as shown in Fig.~\ref{fig:effectiveheatband} and their peak bolometric luminosities (brown lines in Fig.~\ref{fig:bolol}) are also on the low side. 

Another way in which differences in nuclear heating manifest themselves is in the shape of the luminosity curve at late times. 
With lower $Y_e$ simulations, it is more likely that nucleosynthesis will proceed to heavy nuclei that can fission. 
As discussed in Secs. \ref{sec:method} and \ref{sec:source}, this boosts the effective heating rate, both because of the large Q-values involved in fission and also due to the enhanced thermalization of fission products. 
In contrast, the element synthesis in simulation 1 never reaches nuclei which fission, and only $\beta$-decays contribute significantly to the heating.  
Therefore, the luminosity in simulation 1 (dark purple line in Fig.~\ref{fig:bolol}) drops much more steeply at late times. 
We explore the landscape of decaying nuclei more closely in Section~\ref{fission}.

If the effective heating rate is the same, the higher opacity that stems from a larger $X_{\rm Ln+An}$ results in an elongated light curve and a later, dimmer peak, as compared to a simulation with a smaller $X_{\rm Ln+An}$. 
For example, simulations 1 and 12 have similar effective heating rates (dark purple and green lines in Fig.~\ref{fig:effectiveheatband}) at several days, but the peak bolometric luminosity of simulation 1 (dark purple line in Fig.~\ref{fig:bolol}) is noticeably earlier than the peak bolometric luminosity of simulation 12 (dark green line in Fig.~\ref{fig:bolol}). 
This is consistent with the subplot of Fig.~\ref{fig:effectiveheatband}, where simulation 1 (dark green bar) has an $X_{\rm Ln+An}$ which is less than $10\%$ while simulation 12 (dark green bar) has an $X_{\rm Ln+An}$ of about $30\%$.
We caution that while lanthanides and actinides produce a high opacity component, the relationship is one of diminishing returns. 
Increasing the lanthanide mass fraction from 0.1 to 1.0 raises the mean opacity by a factor of less than 2 and even an increase from 0.01 to 1.0 results in a change of a factor of $\lesssim 5$ \citep{Kasen2013Opaci}. 
These differences tend to be less significant than that of the variation in nuclear heating across models. 
If we were working in the context of a fixed heating rate which is insensitive to nuclear physics variations, then the ejecta mass would be tightly constrained by a comparison with observed luminosity, and with this information about the mass, then the opacity could be constrained by matching the evolution of the light curve. 
Since in our study the heating rate is calculated to be consistent with the nuclear model, we can see from Fig.~\ref{fig:bolol} that the (smaller) differences in opacity can be offset by a combination of the heating rate and ejecta mass, or vice versa.

The combination of the effective heating and $X_{\rm Ln+An}$ leads to substantial differences in the duration of the bolometric light curve peaks shown in Fig.~\ref{fig:bolol}. 
Simulations 3 and 4 (blue lines in Fig.~\ref{fig:bolol}) have a fast rise at about 1 day, and their peak luminosity begins to decline only around 12 days. This correlates with the shallow slope in the effective heating rate which comes from an early time fission contribution, as well as a substantial $X_{\rm Ln+An}$, which increases the diffusion time of photons through the ejecta.
These long plateaus are interesting as a potential signal of early time fission and need to be verified with a more sophisticated radiation transport model (to be presented in Barnes \textit{et al.}, in prep).

\subsection{Inferred Ejecta Mass}

Fig. \ref{fig:bolol} demonstrates that the single-component (red) model of a given mass and velocity, as outlined in Section~\ref{sec:method}, could have a wide range of predicted luminosities, depending on the nuclear physics and astrophysics inputs chosen for the \rp{} calculation. 
We now turn to an exploration of the uncertainty in the inferred ejecta mass given uncertainties in these inputs in the case where there is a component dominated by lanthanide/actinide rich material.

\begin{figure}[!h]
  \gridline{
  \fig{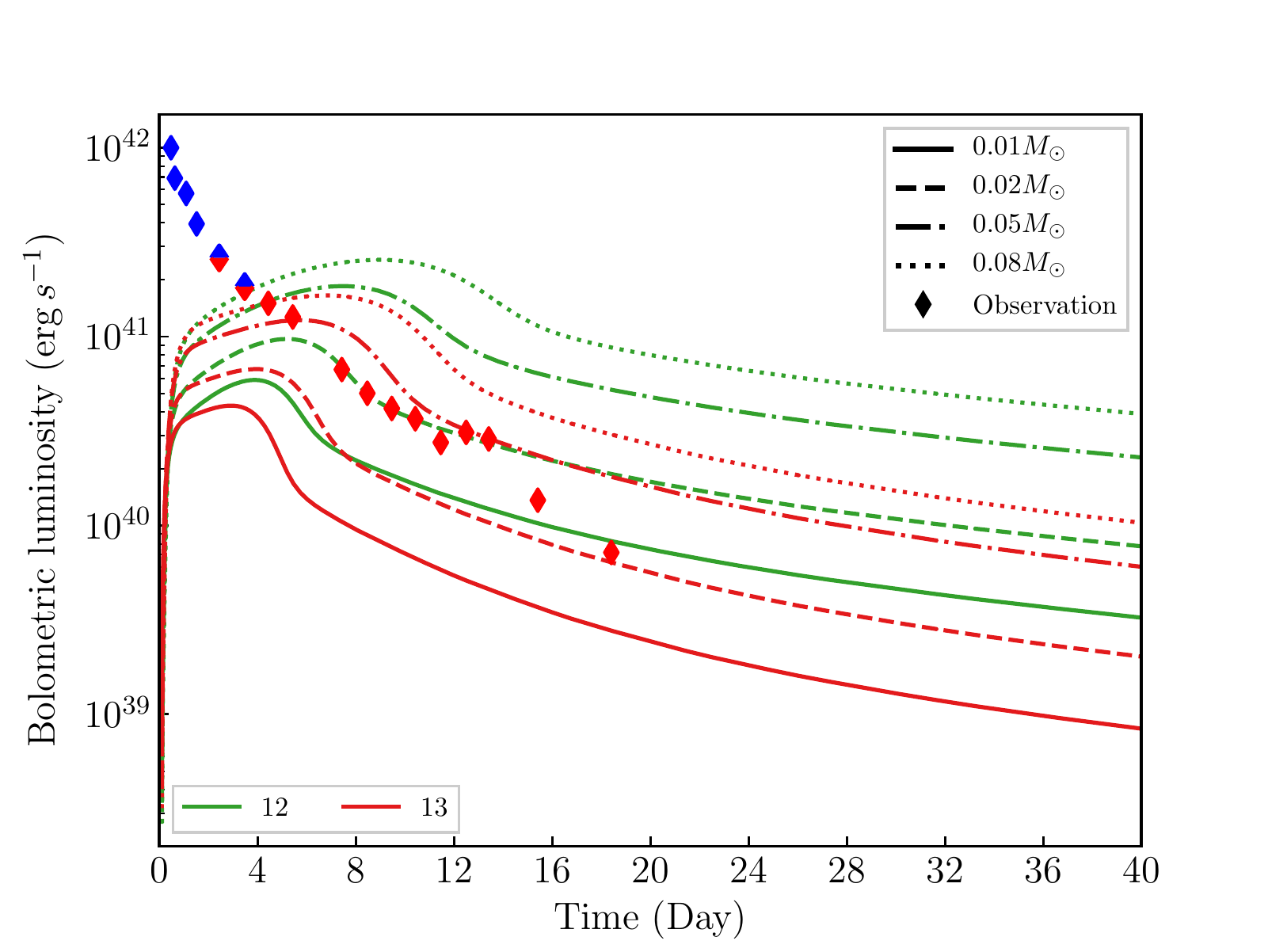}{0.49\textwidth}{a) Bolometric luminosity of the two linear combinations of simulations, labeled as 12 and 13 in Table~\ref{tab:13calc}.}
  {\fig{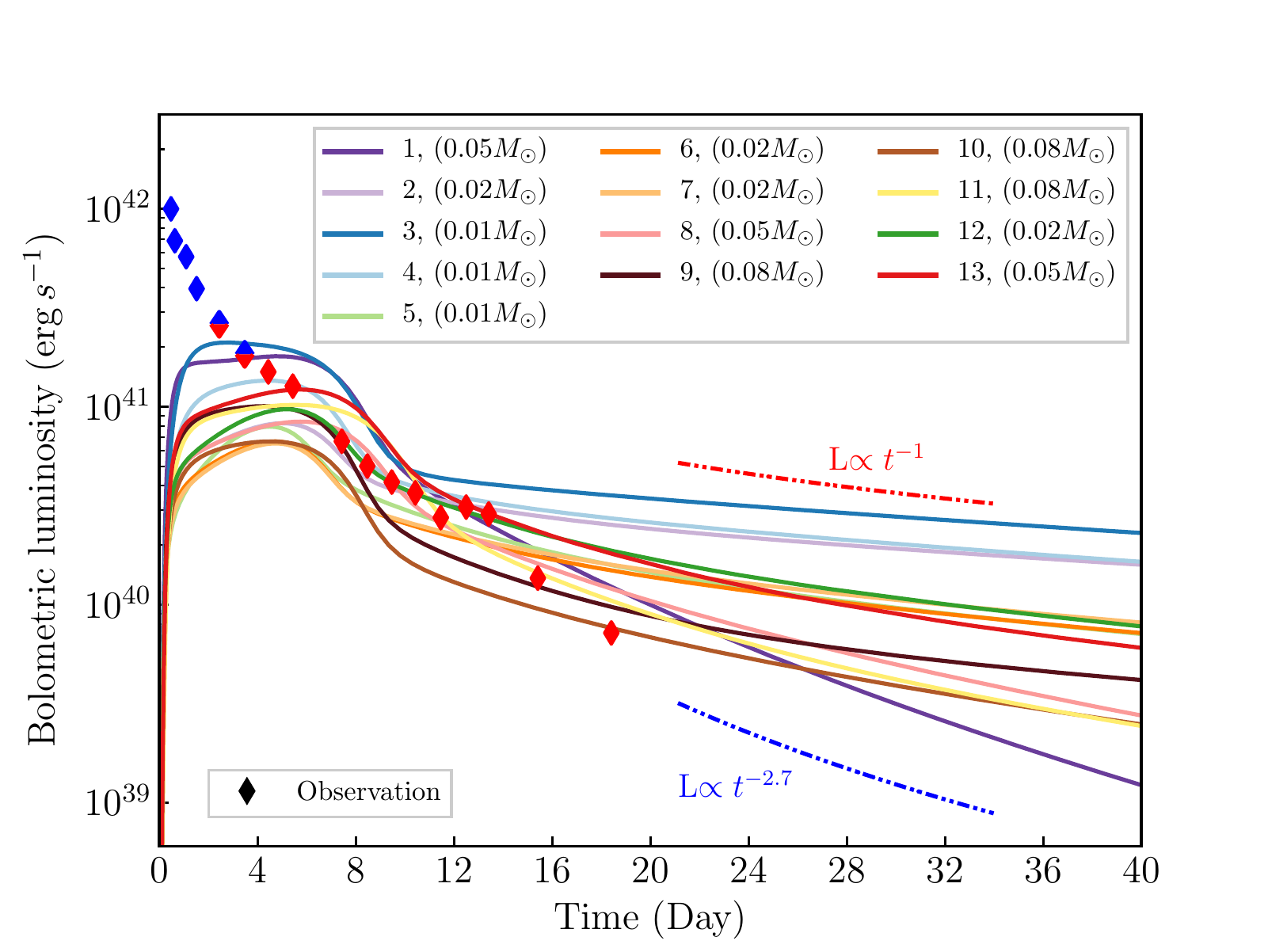}{0.49\textwidth}{b) Bolometric luminosity of 13 simulations with ejecta mass which most closely correspond to the observed bolometric luminosity during $4$ to $12$ days. Lines that are proportional to $t^{-\alpha}$ are shown for comparison.}}}
  \caption{
  \label{fig:fitmassdata} Bolometric luminosity of simulations using different ejecta masses. 
  The ejecta masses are chosen from a grid that includes four ejecta mass parameters: $0.01\,M_{\odot}$ (solid), $0.02\,M_{\odot}$ (dashed), $0.05\,M_{\odot}$ (dash-dot), $0.08\,M_{\odot}$ (dotted), for each  simulations labeled in Table~\ref{tab:13calc}.
  Black diamond markers are observed bolometric luminosity from the KN associated with GW170817~\citep{Drout2017Light}.} 
\end{figure}

In Fig~\ref{fig:fitmassdata}(a), for simulations 12 and 13 (which are the linear combinations that fit the solar abundance pattern), we show the calculated light curve for various ejecta masses ranging from $0.01\,M_{\odot}$ to $0.08\,M_{\odot}$. 
In all models, the average ejecta velocity was $v_{ej}=0.15c$.
As can be seen from Fig.~\ref{fig:fitmassdata}(a), this grid of masses covers a substantial range of bolometric luminosities. 
Additionally, the peak luminosity, width of the peak, and decay of the light curve also depend on the ejecta mass. 
In the context of the methods outlined in Section \ref{sec:method}, if we assume the merger event GW170817 to have synthesized nuclear species in ratios consistent with the solar abundance pattern (for $A > 125$) in conditions like simulations 12 or 13, $0.02 M_{\odot}$ is the best inferred ejecta mass to match the light curve from $8$ to $12$ days.
Additional work with a more complete set of solar abundance-tuned models is needed to determine the full extent of the uncertainty on the bolometric luminosity from models which match the solar abundance pattern but have different nuclear inputs.

Next, we take simulations 1-11 and calculate a light curve for four different values of the ejecta mass:  $0.01\,M_{\odot}$, $0.02\,M_{\odot}$, $0.05\,M_{\odot}$, and $0.08\,M_{\odot}$.  
In Fig.~\ref{fig:fitmassdata}(b), for each simulation, we plot the light curve that best matches the data at $4$ to $12$ days and show the corresponding ejecta mass in the key. 
We see from this figure that a single set of data can be fairly well fit by a variety of ejecta masses.  
The inferred ejecta mass can be relatively small with the simulations with higher effective heating rates and $X_{\rm Ln+An}$. 
For example, the best match for simulations 3, 4, and 5 (HFB and ETFSI mass models) occurs with an ejecta mass of $0.01\,M_{\odot}$. 
The simulations with lower effective heating rates at the time their light curve peaks and lower mass fractions of lanthanides and actinides, such as simulations 9 and 10 with SLY4 mass model, have higher inferred ejecta masses, $0.08\,M_{\odot}$.  
The ejecta mass inferred from simulations 1, 2, 6, 7, 8, 12, and 13, which have light curves that reflect an interplay between the amount of lanthanides and actinides and the effective heating rates, is in between these two extremes. 

This analysis indicates the extent to which the inferred ejecta mass can vary with different nuclear physics and different initial neutron richness, while keeping the rest of the parameters in the KN model constant.  
In particular, it is clear that if the model is not required to be consistent with solar abundance ratios, the inferred ejecta mass may suffer from large uncertainties.
Note that this analysis is intended to show the range of uncertainty that would exist if one considered material that is synthesized with $Y_e < 0.24$ in the 8-12 day range.
Our analysis considers scenarios where the ejecta reaches rather high $X_{\rm Ln+An}$. Previous estimates of an $X_{\rm Ln+An}\sim10^{-2}$ (e.g. \citet{Chornock2017TheEl} calculated based on observations are derived under specific assumptions about the \rp{} heating rate, which we vary. 
If a smaller mass fraction is preferred than is obtained from neutron-rich element synthesis, then these curves could be combined with lanthanide-free material to reach the desired lanthanide mass fraction, and the previous analysis could be repeated along with complete fits to the spectra and time evolution of the luminosity. 

\subsection{Power Law Fits for Late-Time KN signals}
For some practical applications, it may be useful to approximate the light curve as a power law. For the power law $t^{-\alpha}$ we find this range to be $1 \leq \alpha \leq 2.7$, depending on the nuclear physics inputs, for the time period we consider, $t < 40$ days.
The broad range of power-law indices we find reflects the fact that our analysis includes contributions from all decay channels, including $\alpha$-decay and fission, which can contribute significantly to the heating. 
For example, Simulation 1 (purple line in Fig.~\ref{fig:fitmassdata}(b)), is not neutron rich enough to significantly produce heavy nuclei which undergo spontaneous fission or $\alpha$-decay. Therefore the late-time decay is not supplemented by these extra processes, and the bolometric luminosity decays rapidly, proportional to $t^{-2.7}$. Meanwhile, simulations 3 to 8 have more significant fissioning and/or $\alpha$-decaying nuclei, and show a much slower decay rate that is closer to $t^{-1}$.

Other estimates of this power law at a different times can be found in \citet{Kasen2019Radioa,Waxman2019Latet,Hotokezaka2020Radioa}.

\section{Important Reactions for Nuclear Heating}\label{fission}

In this section, we explore variations and similarities in the predictions of individually contributing isotopes in each decay channel, and how these account for some of the differences in overall heating explored in previous sections. 
In order to obtain a reasonably full picture of important reactions, for the majority of this section, we return to the nearly full suite of simulations combining the 8 theoretical nuclear models and 4 combinations of spontaneous fission prescriptions and daughter product distributions as described in Section~\ref{method:nucleosyn}, yielding 32 combinations for each of the $Y_{e} \leq 0.24$ values (simulations with $Y_{e}=0.28$ showed little to no fission past one day) described in Section~\ref{method:astro} for a total of 224 simulations.

\begin{figure}[!h]
  \centering
\includegraphics[angle=0,width = 0.69\textwidth]{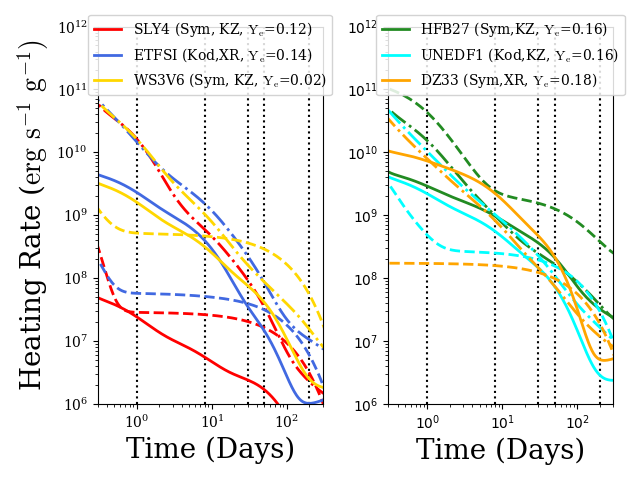}
\caption{\label{fig:indchannels} Heating from $\alpha$-decay (solid), $\beta$-decay (dash-dotted) and spontaneous fission (dashed) for six different simulations, selected to highlight the competition between respective channels at different times. Times of 1, 8, 30, 50 and 200 days are indicated with vertical lines. 
The parenthesis in the legend indicate the spontaneous fission yield, spontaneous fission rate, and $Y_{e}$, in that order.}
\end{figure}

While in previous sections we examined the fractional effective heating in each channel, we instead show in Fig.~\ref{fig:indchannels} the absolute $\alpha$-decay, $\beta$-decay and spontaneous fission heating rates as a function of time for a selection of simulations. 
These were selected with the intention of sampling from the range of $Y_{e}$, nuclear models, and spontaneous fission prescriptions, while capturing some of the interesting behavior. 
We find general agreement across all simulations that the $\beta$-decay heating follows a power law decay, as illustrated by the dot-dashed lines for the six simulations in this figure. 
When we compare all simulations with $Y_{e} \leq 0.21$, we find that the absolute $\beta$-heating varies by approximately one order of magnitude, increasing to up to two orders of magnitude at late times ($\sim100$ days).
However, $\beta$-decay is not always the dominant reaction channel, and Fig.~\ref{fig:fractionss1624} serves as an example of how there are a range of possibilities for the dominating channels at essentially all times.
In contrast to the relative similarity in $\beta$-decay heating across models, $\alpha$-decay heating (solid lines in Fig.~\ref{fig:indchannels}) varies more in the overall amount of heating by between two to four orders of magnitude. 
The spontaneous fission heating rate (dashed lines in Fig.~\ref{fig:indchannels}) shows the most sensitivity to model inputs, displaying differences in overall shape as a function of time and between three to six orders of magnitude variation in its value.
We note that in the cases of $\alpha$-decay and spontaneous fission, some simulations with $Y_{e} = 0.24$ and $Y_{e} = 0.28$ show no heating at the times included in Fig.~\ref{fig:indchannels}. Therefore, including these high $Y_{e}$ cases would amplify these uncertainties by several orders of magnitude by extending them to zero.

\subsection{Spontaneous Fission\label{sec:spfheating}}

\begin{figure}[!ht]
\gridline{
\fig{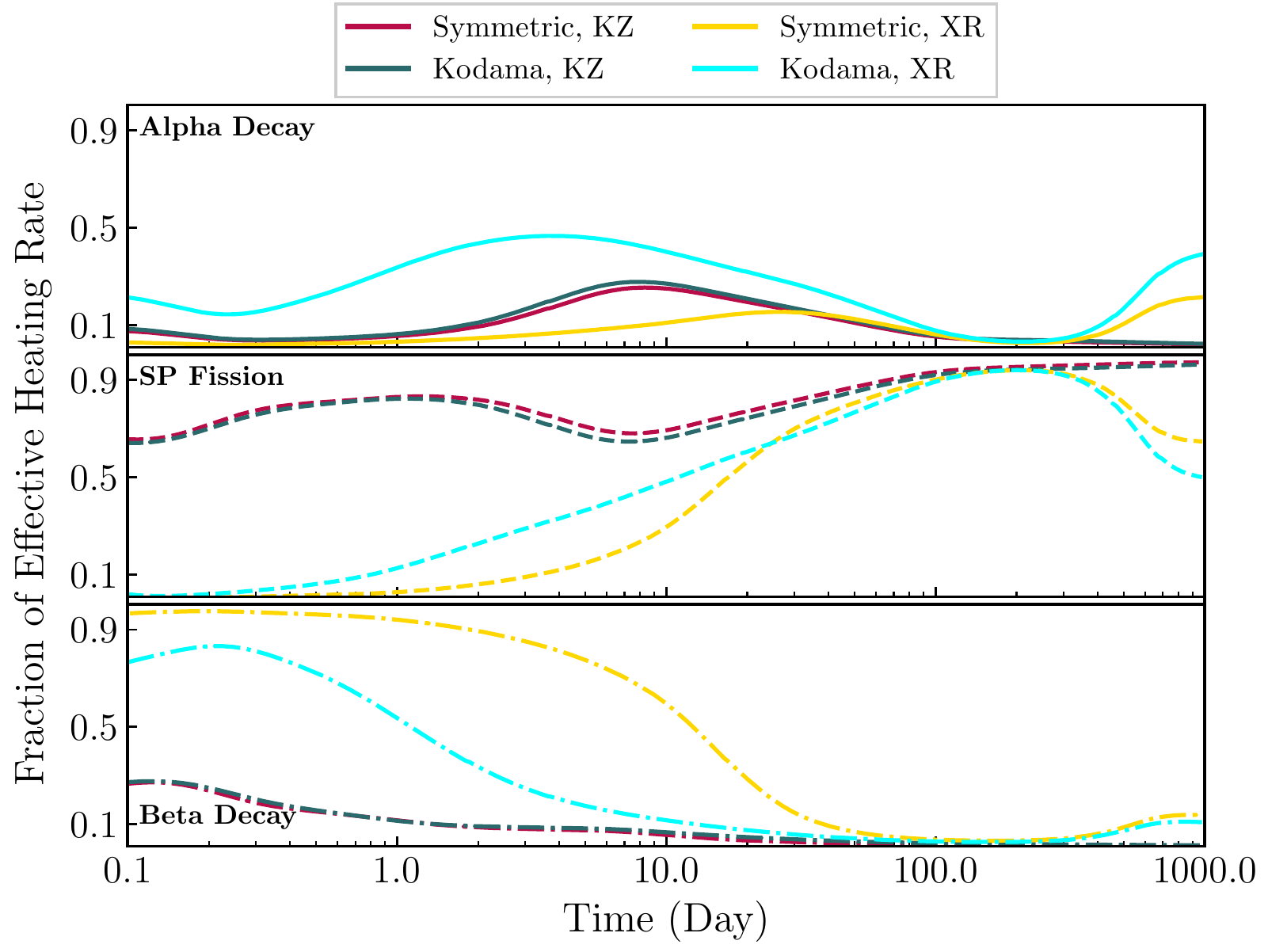}{0.49\textwidth}{(a) Simulations using HFB27 theoretical nuclear model and with electron fraction $Y_e=0.16$.}             
\fig{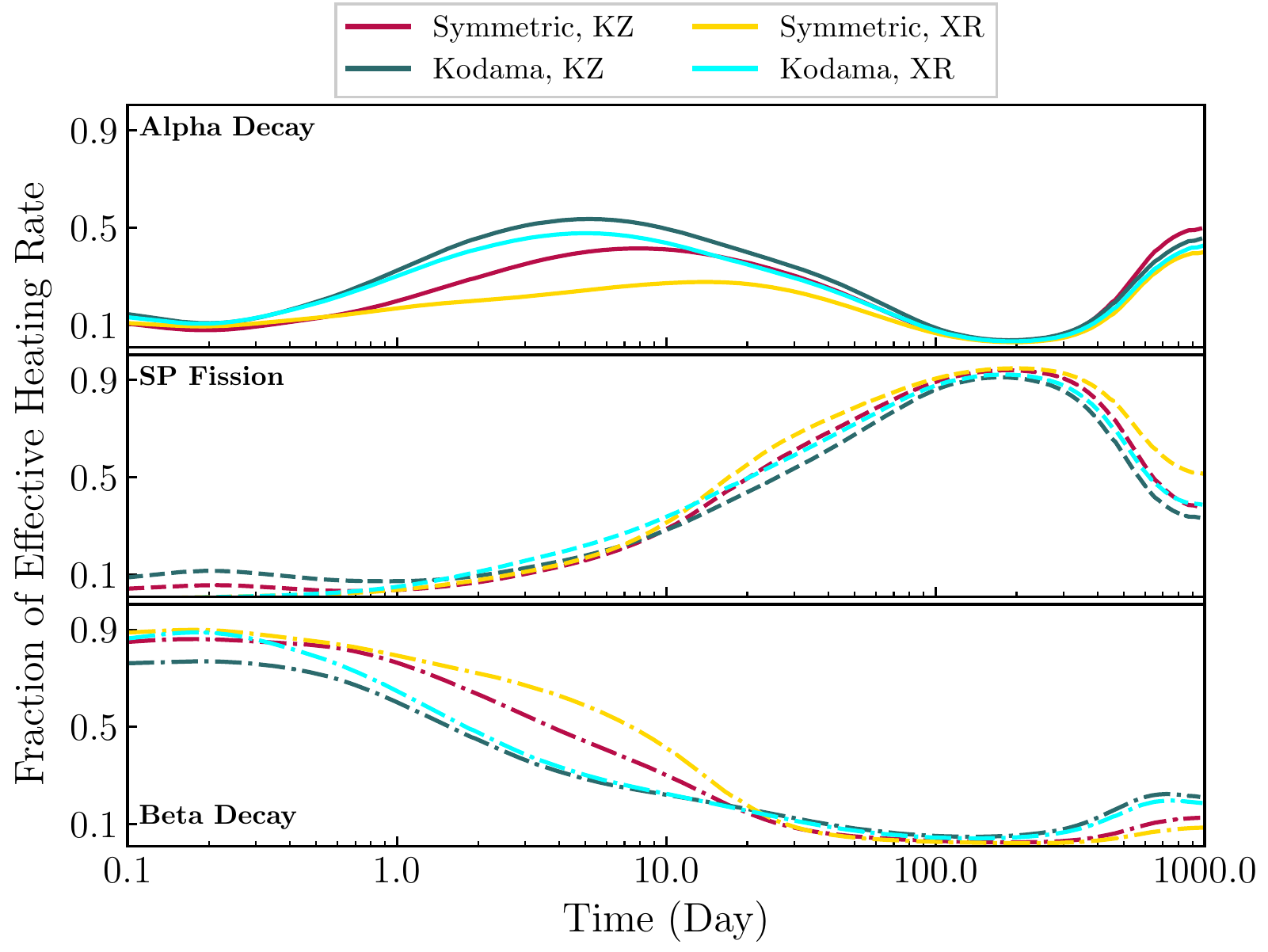}{0.49\textwidth}{(b) Simulations using UNEDF1 theoretical nuclear model and with electron fraction $Y_e=0.16$}}
\caption{\label{fig:fractionfis}Comparison of the impact of different spontaneous fission prescriptions for two different nuclear models (left: HFB27, right: UNEDF1) on the fractional effective heating rates.}
\end{figure}

Given the potential importance of spontaneous fission, as well as the large uncertainties associated with spontaneous fission heating, we examine the consequences of spontaneous fission rate prescriptions as well as the influence of daughter product distributions.  
In Fig.~\ref{fig:fractionfis} we show fractional effective heating rates in order to compare simulations for which we altered only these aspects of spontaneous fission, while holding all other nuclear and astrophysical inputs constant.
The most striking takeaway from Fig.~\ref{fig:fractionfis}(a) is that prior to tens of days, there can be a clear difference between the results with KZ and XR for cases which permit nuclei with high mass numbers to be significantly populated, as is the case for HFB with KZ rates.
This is because XR spontaneous fission rates are relatively high starting around $Z\sim94$ regardless of the predicted fission barriers since this prescription has no dependence on this input. Therefore in the XR case, high mass, high fission Q-value nuclei are never significantly populated.
We see also from both panels of this figure that changing the daughter product distribution can influence the fraction of $\beta$-decay heating.  

\begin{figure}[!h]
\gridline{
\fig{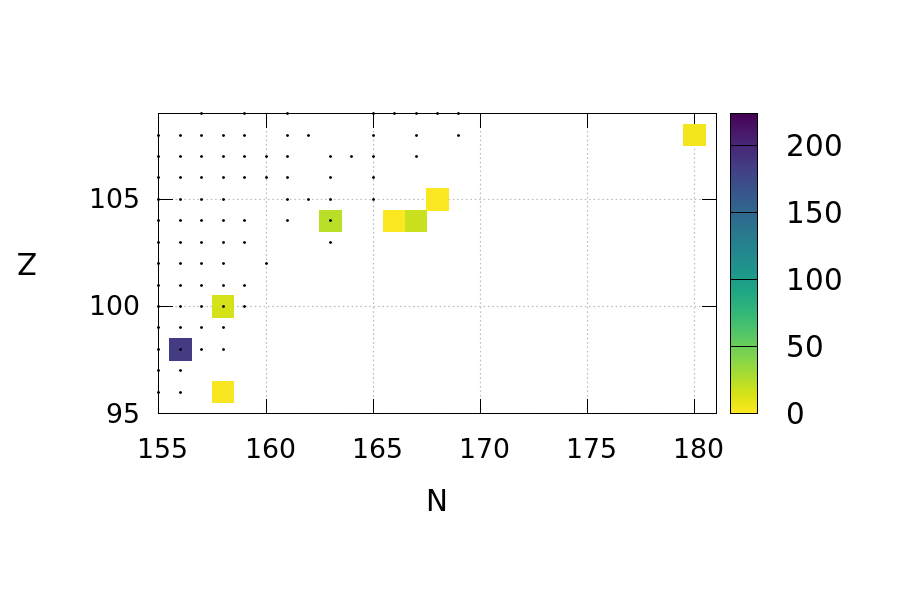}{0.33\textwidth}{(a) Spontaneous fission at 1 day.}
\fig{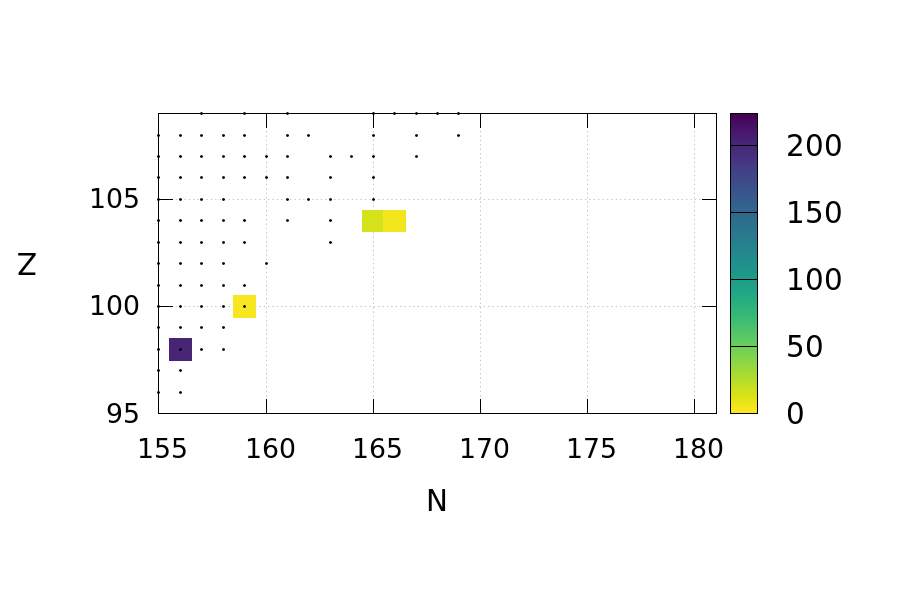}{0.33\textwidth}{(b) Spontaneous fission at 8 days.}
\fig{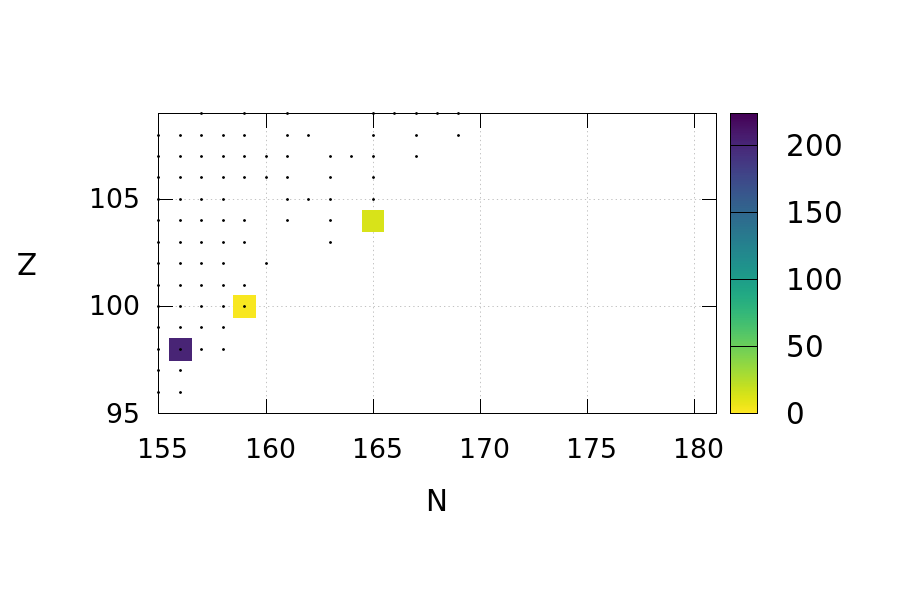}{0.33\textwidth}{(c) Spontaneous fission at 50 days.}}
\caption{Isotopes responsible for 80\% of the total spontaneous fission heating rate. Colors indicate the number of times (out of 224 simulations with different nuclear physics and astrophysical inputs) each isotope was listed as being part of the top 80\% of spontaneous fission heating. 
Dots indicate nuclei which have measured values for their spontaneous fission rate. \label{fig:spfisotopes} }
\end{figure}

Returning to all 224 simulations, we now identify those isotopes responsible for most of the spontaneous fission heating. 
In general, we determine key contributors at each selected time by ranking the heaters within each channel.  We then sum this list until 80\% of the total heating in that channel is accounted for, as we find that 80\% adequately represents the diversity of significant contributors. 
In Fig.~\ref{fig:spfisotopes}, we show a section of the chart of the nuclides, graphically representing the number of simulations for which each isotope was responsible for all or part of the top 80\% of the spontaneous fission heating. 
We find that in most cases, even at early times, \Cf~($t_{1/2}=60.5\pm0.2$ days)~\citep{cf_measure} is among the top isotopes contributing to spontaneous fission heating. 
This is reflected in the behavior of the spontaneous fission heating lines in Fig.~\ref{fig:indchannels}, showing a consistent underlying plateau and decay after about 50 days, indicating a population and evolution of \Cf~on its spontaneous fission timescale. 
However, in some cases, we find other isotopes to be responsible for all or part of 80\% of the spontaneous fission heating in that model, either because they dominate over \Cf~or because \Cf~itself does not make up the entire 80\%. 
We list the simulations in which an isotope {\it other} than \Cf~is responsible for all or part of $80\%$ of the total spontaneous fission heating in Table~\ref{isospf}. 
The dominant trend is that models which permit the synthesis of heavy species near or beyond the N=184 predicted shell closure have the greatest diversity in nuclear species that contribute to spontaneous fission heating, as is the case with HFB + KZ combinations.
We note, additionally, that very low fission barriers will result in limited population of nuclei near \Cf~ \citep{Vassh2019Using,giuliani2019fission}, and thus little to no contribution of spontaneous fission to the nuclear heating.
At times much earlier than the timescale for spontaneous fission of \Cf, e.g. at one day, our simulations have a more diverse set of important spontaneous fission heaters than at later times, especially from nuclei which decay on shorter timescales.

 \startlongtable
 \begin{deluxetable}{ccllll}
 \tablecaption{\label{isospf}Top Contributing Nuclei: Spontaneous Fission (80\%)} 
 \tablehead{\colhead{Time (d)} & \colhead{Isotope} & \colhead{Model} & \colhead{Fission Yield} & \colhead{Fission Rate} & \colhead{$Y_{e}$}}
 \startdata
 $1$ & \ce{^{254}_{96}Cm} & HFB22 & Symmetric & XR & 0.24*\\
     &   &   & Kodama & XR & 0.24*\\
     & \ce{^{258}_{100}Fm} & HFB27 & Symmetric & KZ & (0.02-0.24)*\\
     &   &   & Kodama & KZ & (0.02-0.24)*\\
     & \ce{^{267}_{104}Rf} & FRDM2012 & Kodama & KZ & 0.12, 0.14\\
     &   & UNEDF1 & Symmetric & KZ & 0.14, 0.18\\
     &   &   & Kodama & KZ & (0.02-0.18)\\
     &   & HFB22 & Symmetric & KZ & 0.02*, 0.12, 0.14, 0.16*, 0.18*, 0.21, 0.24\\
     &   &   & Kodama & KZ & (0.02-0.18)*, 0.21, 0.24\\
     & \ce{^{270}_{104}Rf} & HFB22 & Symmetric & KZ & 0.02*\\
     & \ce{^{271}_{104}Rf} & HFB22 & Symmetric & KZ & 0.02*, 0.12, 0.14, 0.16*\\
     &   &   & Kodama & KZ & (0.02-0.16)*\\
     &   & HFB27 & Symmetric & KZ & (0.02-0.18)*\\
     &   &   & Kodama & KZ & (0.02-0.18)*\\
     & \ce{^{273}_{105}Db} & UNEDF1 & Kodama & KZ & 0.12\\
     & \ce{^{288}_{108}Hs} & ETFSI & Symmetric & KZ & 0.02, 0.12, 0.14*\\
     &   &  & Kodama & KZ & 0.14\\
 $8$ & \ce{^{259}_{100}Fm} & HFB22 & Symmetric & KZ & 0.18\\
     &   &  & Kodama & KZ & 0.18\\
     & \ce{^{269}_{104}Rf} & HFB22 & Symmetric & KZ & 0.16, 0.18\\
     &   &  & Kodama & KZ & 0.02, (0.14-0.18)\\
     &   & HFB27 & Symmetric & KZ & 0.02, 0.12, 0.16, 0.18\\
     &   &  & Kodama & KZ & 0.02, 0.12, 0.16, 0.18\\
     & \ce{^{270}_{104}Rf} & HFB27 & Symmetric & KZ & 0.02, 0.18\\
     &   &  & Kodama & KZ & 0.02, 0.14\\
 $50$& \ce{^{259}_{100}Fm} & HFB22 & Symmetric & KZ & 0.18\\
     &   &  & Kodama & KZ & 0.18\\
     & \ce{^{269}_{104}Rf} & HFB22 & Symmetric & KZ & 0.16\\
     &   &  & Kodama & KZ & 0.02, 0.16, 0.18\\
     &   & HFB27 & Symmetric & KZ & 0.02, 0.12, 0.16, 0.18\\
     &   &  & Kodama & KZ & 0.02, 0.12, 0.14, 0.16, 0.18\\
\enddata
\tablecomments{Simulations in which an isotope other than \Cf~was responsible for all or part of 80\% of the total spontaneous fission heating. 
The right columns list out the specific simulations in which each isotope was seen, labeled first by nuclear model, then in parentheses by spontaneous fission yield, spontaneous fission rate and $Y_{e}$. 
An asterisk indicates that for the corresponding simulations, \Cf~did \emph{not} appear in the list of top 80\%.}
\end{deluxetable}

\subsection{Key Reactions in Alpha-Decay}
\begin{figure}[h!]
\gridline{
\fig{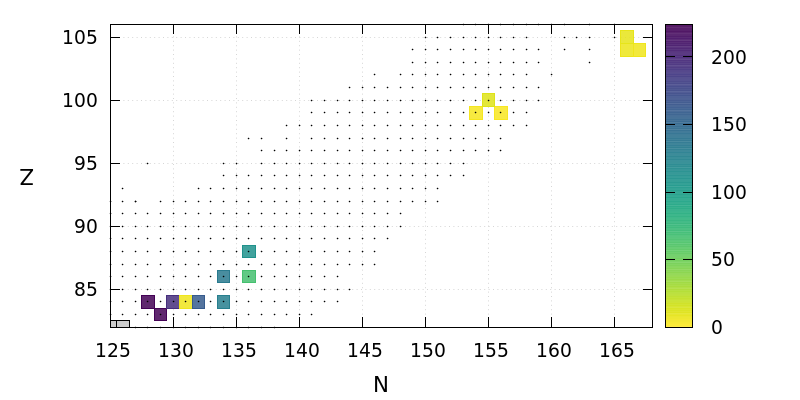}{0.33\textwidth}{(a) $\alpha$-Decay at 1 day.}
\fig{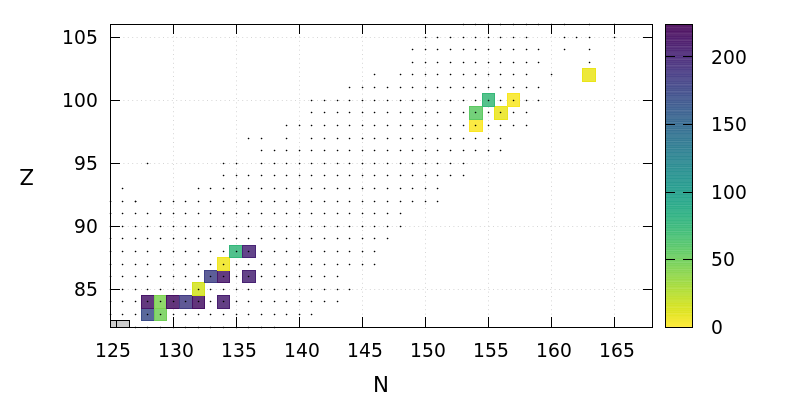}{0.33\textwidth}{(b) $\alpha$-Decay at 8 days.}
\fig{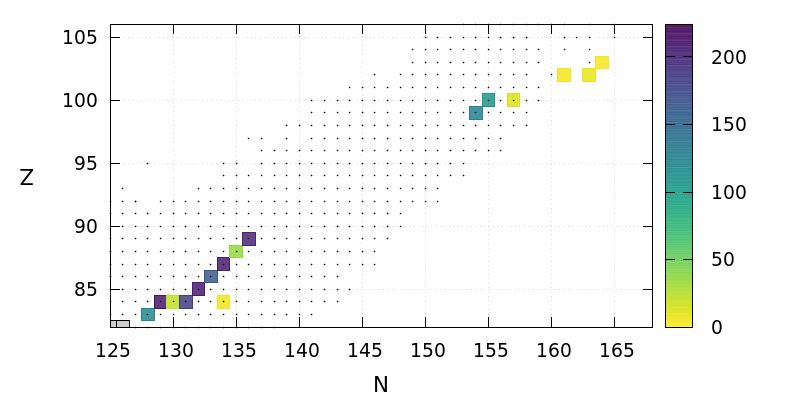}{0.33\textwidth}{(c) $\alpha$-Decay at 30 days.}}
\caption{\label{fig:adisotopes}Isotopes responsible for 80\% of total $\alpha$-decay heating at indicated times. Colors indicate the number of times (out of 224 simulations with different nuclear physics and astrophysical inputs) each isotope was listed as being part of the top 80\%. 
Dots represent the nuclei for which $\alpha$ decay rates have been measured.}
\end{figure}

As Fig.~\ref{fig:fractionfis} shows, varying the nuclear inputs, such as the mass model or the spontaneous fission rate and yield distribution, can also impact the $\alpha$-decay heating prediction.
Depending on the specific model inputs, we find that $\alpha$-decay heating can range from 10\% of the total heating to up to 50\% prior to 100 days.
Following the same procedure described in Section~\ref{sec:spfheating}, we identify isotopes responsible for the top 80\% of $\alpha$-decay heating.
In Fig.~\ref{fig:adisotopes}, we show the evolution of the top contributing nuclei between 1, 8, and 30 days, with a more complete list included in the appendix. 
At each of the selected times, there is generally good agreement across simulations regarding which isotopes contribute the most.

We distinguish two regions of importance: $83\le Z \le89$ and $98\le Z \le105$. 
The first region, shown in the bottom left corner of each subplot of Fig.~\ref{fig:adisotopes}, is accessible to most trajectories, and is a region where the preferential decay mode tends to be $\alpha$-decay.  
The magnitude of the $\alpha$-heating in this region is largely determined by the amount of material which reaches these high mass regions, thus we find that it exhibits a relatively high sensitivity to $Y_{e}$. 
Comparing the important isotopes in this region as a function of time (across panels), we see the expected evolution toward stability. 
 
Meanwhile, in the upper right region of the nuclear chart (higher atomic numbers), decay modes display a stronger competition between $\alpha$-decay and spontaneous fission, e.g. one species might primarily decay via $\alpha$-decay with a secondary or tertiary spontaneous fission decay, while its immediate neighbor might primarily decay via spontaneous fission. 
Therefore, while the $\alpha$-heating in this region is sensitive to the amount of material which reaches the region, it is also sensitive to the relative rates of $\alpha$-decay and spontaneous fission along the relevant decay chains. 
By examining the ordered list of contributors to $\alpha$-decay heating, we find that in simulations where isotopes from this second region are important, they tend to supply an increasing fraction of the total $\alpha$-decay heating as time progresses. We find that the two most consistent isotopes from the second region, \isotope[253]{Es} and \isotope[255]{Fm}, are present in most nuclear model simulations from this sample, but are restricted to simulations using lower $Y_{e}$ values (only below 0.18).
Finally, while many nuclei in this region have measured $\alpha$-decay rates, some do not, making both theoretical predictions and measurements for these $\alpha$-decays particularly important.

\subsection{Impact on Beta-Decay Heating}

\begin{figure}[h!]
\centering
\includegraphics[angle=0,width = 1.0\textwidth]{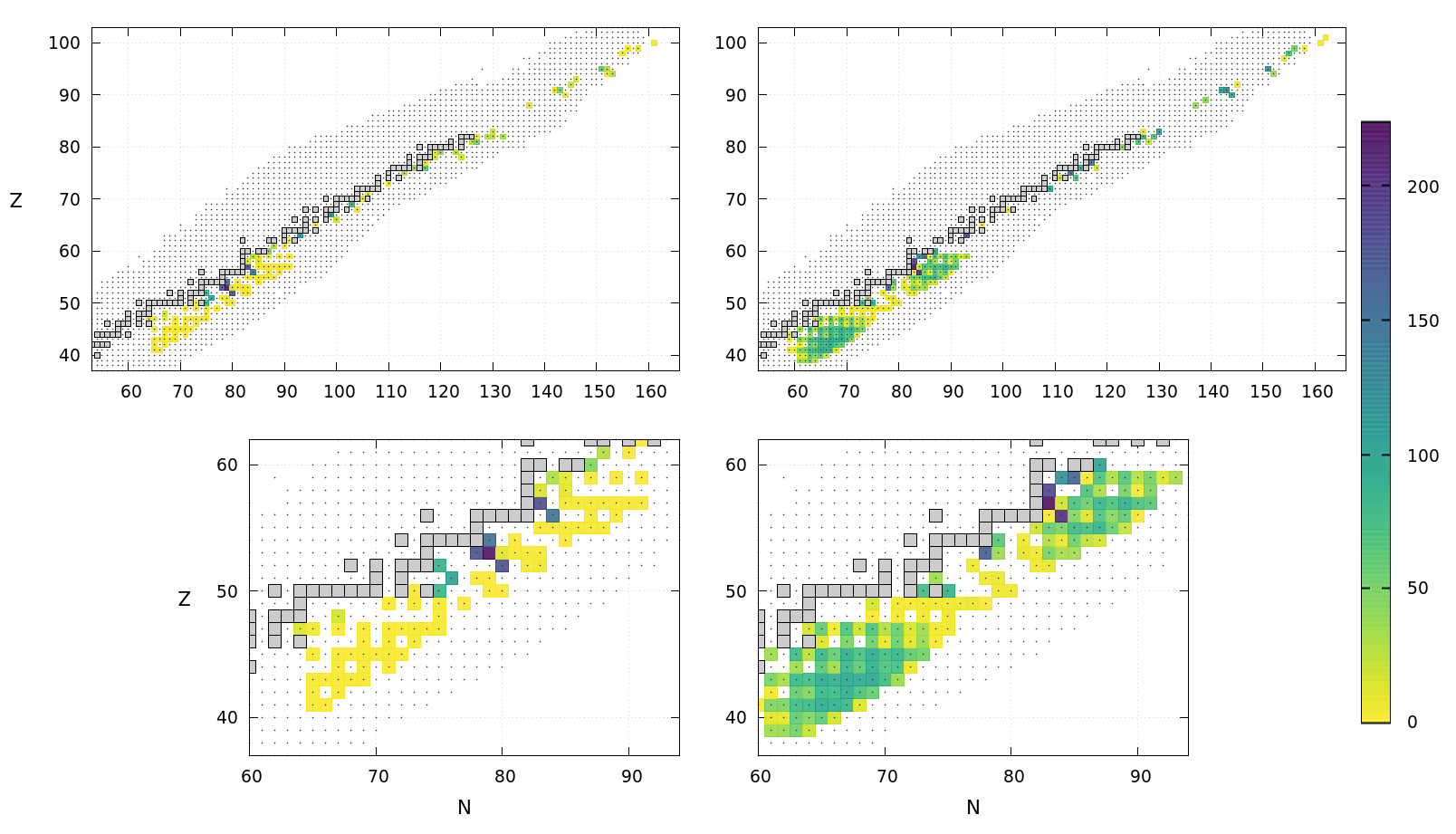}
\caption{Top panel: Frequency of $\beta$-decay nuclei in the top 80\% of the total $\beta$-heating rate at 8 (left) and 50 (right) days. 
Colored bar indicates the number of simulations (out of 224 total) in which each isotope was listed as being part of the top 80\%. 
Dots indicate isotopes which have a measured beta decay rate. 
Bottom panel: Zoom on the region where more consistent spontaneous fission daughter products appear after at least 50 days.} 
\label{fig:kfigbeta}
\end{figure} 

We find the largest variation in fractional effective heating in the contribution from $\beta$-decay heating. 
Depending on the time post-merger, varying the fission prescription, nuclear models and astrophysical conditions can result in $\beta$-decay heating comprising anywhere between 10\% to more than 90\% of the total effective heating.
The spontaneous fission reactions and $\alpha$-decays discussed in the previous sections have direct impacts on the $\beta$-decay heating, as these processes are a significant factor in determining the population of nuclei which eventually $\beta$-decay. 
Fig.~\ref{fig:kfigbeta} shows the frequency of isotopes responsible for all or part of 80\% of the $\beta$-decay heating at 8 (left) and 50 (right) days: a more complete list is found in the appendix. 
Close to stability, there are a few important isotopes that show up in nearly every simulation.
Further away from stability, and as early as 30 days, we start to see the significant impact of \Cf~spontaneous fission daughter products in the $A=130$ region. 
Given that the simulations we compare in this section span $Y_{e}$ values from 0.02 to 0.24, it is expected that the abundance of \Cf~varies from simulation to simulation, and therefore the fraction of $\beta$-heating which comes from decays of daughter products of \Cf~is also not uniform; this is reflected in Fig.~\ref{fig:kfigbeta}. 
Surrounding the two distinct green patches seen in the bottom right panel of Fig.~\ref{fig:kfigbeta} is a more diffuse distribution of isotopes (yellow patches) that appear to be significant in few simulations. 
These are primarily found in simulations using HFB nuclear models with Kodama fission yields and KZ spontaneous fission rates.
This highlights the potential importance of isotopes with unmeasured characteristics in the high-Z region ($100\le Z\le106$), both directly via spontaneous fission or $\alpha$-decay, as well as indirectly via the decay products of these processes. 
The amount of heating from spontaneous fission, $\alpha$-decay, and subsequently, $\beta$-decay, is ultimately sensitive to whether  the synthesis of heavy nuclei reaches this high-Z region, and if it does, the process by which they preferentially decay.

\section{Conclusion}
Observations of the electromagnetic signal that comes from a NSM present a unique opportunity to investigate both the type and amount of the elements that are ejected in these events. 
It is generally believed that a high opacity material, likely rich in lanthanides, was ejected from the recent NSM GW170817.  
These elements are synthesized via the \rp{}, so this suggests that at least some \rp{} material was produced in this NSM. 

Using the observed electromagnetic signal to infer elements produced in the merger is an inverse problem. 
Since uncertainties are present in a variety of inputs relevant for predicting light curves, interpreting merger signals is subject to degeneracies. 
From the results of our investigations, we estimate that a range of approximately one order of magnitude in the ejected mass is needed to reconcile light curve observations with the broad range of nuclear physics uncertainties we have explored.
Our estimate by design includes simulations with a variety of final abundance patterns. 
While the two simulations we considered that produce a solar main \rp{} seem to show more similar trends, we caution that a more extensive study of nuclear physics models is needed to more quantitatively estimate the full range of behavior when applying a solar pattern constraint. 
However, it is too preliminary to impose such a constraint without definitive evidence that NSMs are the sole site of \rp{} element production.

We find that the late time behavior of the light curve is substantially different above and below $Y_e \sim 0.21$ for our choice of outflow timescale and entropy. 
We focus on below $Y_e \sim 0.24$ conditions, and find that the effective heating rate exhibits much more variability than the opacity.
Although lanthanides can be produced with $Y_{e}\lesssim 0.24$, there is very little actinide production above $Y_{e}\sim0.21$, resulting in less spontaneous fission, and a rapidly decaying bolometric luminosity at around ten days. 
Meanwhile for $Y_{e}\lesssim0.21$, considerably more spontaneous fission occurs and the bolometric luminosity decays much more slowly.  
This is directly relevant to the open question of dominant \rp{} production sites within the merger, including dynamical ejecta (tidally ejected or \lq\lq squeezed") and winds (from a hypermassive NS or accretion disk).
Although each site is thought to be dominated by its own range of $Y_{e}$, there is variety in the predictions of these ranges and therefore in the nucleosynthetic outcomes within each site. 
This includes a possible admixture with higher $Y_e$ material. 
This points to the critical need for dynamical and wind ejecta models which carefully consider microphysical processes in order to translate observations into an understanding of NSM nucleosynthesis, as well as that of BHNS mergers.

Our results show that single spontaneous fission reactions can have leverage on the bolometric luminosity on both short and long timescales. 
We explore different fission barriers and find that some choices show substantially elevated radioactive heating around one day, which is a key time for setting the light curve's peak luminosity as well as the peak's duration. 
As part of explicitly examining the impact of different fission treatments, we identify key nuclear reactions and their effect on the heating. 
For practical use in light curve models, we include tables of key radioactive decay reactions from our survey and suggest a range of effective heating and bolometric luminosity decay rates. 

Looking forward, our results highlight several needs.  
More detailed studies with sophisticated radiation transport coupled to hydrodynamics are essential; this will anchor the uncertainties reported here to state of the art light curve predictions. 
Examinations of the relationship of the observed light curve to alpha decay and fission contributions can be combined with existing techniques which estimate outflow velocity using absorption features in the observed spectrum, since both methods shed light on the type of outflow from which \rp{} material originates. 
To reduce the uncertainties which stem from nuclear inputs, a concerted effort in both nuclear experiment and theory is needed in all areas relevant to \rp{} nucleosynthesis including nuclear masses, fission rates, fission daughter product distributions, as well as $\beta$-decay and $\alpha$-decay rates.

\section{Acknowledgements}
The work of Y-L.Z., K.L., N.V., G.C.M., M.R.M., and R.S. was partly supported by the Fission In R-process Elements (FIRE) topical collaboration in nuclear theory, funded by the U.S. Department of Energy. 
Additional support was provided by the U.S. Department of Energy through contract numbers DE-FG02-02ER41216 (G.C.M), DE-FG02-95-ER40934 (R.S. and T.M.S.), and DE-SC0018232 (SciDAC TEAMS collaboration, R.S. and T.M.S). 
J.B. is supported by the National Aeronautics and Space Administration (NASA) through the Einstein Fellowship Program,  grant number PF7-180162.
R.S. and G.C.M also acknowledge support by the National Science Foundation Hub (N3AS) Grant No. PHY-1630782. M.R.M. was supported by the US Department of Energy through the Los Alamos National Laboratory. 
Los Alamos National Laboratory is operated by Triad National Security, LLC, for the National Nuclear Security Administration of U.S. Department of
Energy (Contract No. 89233218CNA000001). 
This work was partially enabled by the National Science Foundation under Grant No. PHY-1430152 (JINA Center for the Evolution of the Elements).
The work of K.L. was supported partially through EUSTIPEN (Europe-U.S. Theory Institute for Physics with Exotic Nuclei), which is supported by FRIB Theory Alliance under DOE grant number DE-SC0013617. This paper is approved for universal release, assigned LA-UR-20-22630.

\bibliography{ref}

\appendix
We list the key radioactive decay reactions that contribute to the heating from the $\alpha$-decay and $\beta$-decay channels.
We consider eight theoretical nuclear models, two fission yield distributions and two spontaneous fission rate choices, giving 32 combinations of nuclear physics inputs. 
We list the reactions that appear in the top 80\% (as defined in Section \ref{fission}) more than 16 times (black) and more than 28 times (red) out of the 32 combinations for each of the initial electron fractions in Tables 4 and 5. 
In the case of $\alpha$-decays, we find that when present in the ranked list of top contributors, the isotopes listed in Table~\ref{tabad} account for $93\%$ of the top $80\%$ of the $\alpha$-decay heating, on average. Similarly, in the case of $\beta$-decays, we find the isotopes listed in Table~\ref{tabbmd} also account for $93\%$ of the top $80\%$ of the $\beta$-decay heating, on average.

\startlongtable
\begin{deluxetable}{ccc}
\tablecaption{\label{tabad}Key contributing $\alpha$-decays}
\tablehead{\colhead{Time (d)} & \colhead{$Y_e$} & \colhead{$(Z,N)$}}
\startdata
\renewcommand{\arraystretch}{1.5}
 $1$ & $0.02$ & \textcolor{red}{\ce{^{212}_{83}Bi}, \ce{^{212}_{84}Po}, \ce{^{214}_{84}Po}}, \ce{^{216}_{84}Po}, \ce{^{220}_{86}Rn}, \ce{^{224}_{88}Ra}\\ 
     & $0.12$ & \textcolor{red}{\ce{^{212}_{83}Bi}, \ce{^{212}_{84}Po}, \ce{^{214}_{84}Po}}, \ce{^{216}_{84}Po}, \ce{^{218}_{84}Po}, \ce{^{220}_{86}Rn}, \ce{^{224}_{88}Ra}\\
     & $0.14$ & \textcolor{red}{\ce{^{212}_{83}Bi}, \ce{^{212}_{84}Po}}, \ce{^{214}_{84}Po}, \ce{^{216}_{84}Po}, \ce{^{220}_{86}Rn}\\
     & $0.16$ & \textcolor{red}{\ce{^{212}_{83}Bi}, \ce{^{212}_{84}Po}, \ce{^{214}_{84}Po}}, \ce{^{216}_{84}Po}, \ce{^{218}_{84}Po}, \ce{^{220}_{86}Rn}, \ce{^{224}_{88}Ra}\\
     & $0.18$ & \textcolor{red}{\ce{^{212}_{83}Bi}, \ce{^{212}_{84}Po}, \ce{^{214}_{84}Po}}, \ce{^{216}_{84}Po}, \ce{^{218}_{84}Po}, \ce{^{220}_{86}Rn}, \ce{^{224}_{88}Ra}\\
     & $0.21$ & \textcolor{red}{\ce{^{212}_{83}Bi}, \ce{^{212}_{84}Po}, \ce{^{214}_{84}Po}}, \ce{^{216}_{84}Po}, \ce{^{218}_{84}Po}, \ce{^{220}_{86}Rn}, \ce{^{224}_{88}Ra}\\
     & $0.24$ & \textcolor{red}{\ce{^{212}_{83}Bi}, \ce{^{212}_{84}Po}}, \ce{^{214}_{84}Po}, \ce{^{216}_{84}Po}, \ce{^{218}_{84}Po}, \ce{^{220}_{86}Rn}\\
     & $0.28$ & \ce{^{212}_{83}Bi}, \ce{^{212}_{84}Po}, \ce{^{216}_{84}Po}, \ce{^{220}_{86}Rn}, \ce{^{224}_{88}Ra}\\
 $8$ & $0.02$ & \textcolor{red}{\ce{^{212}_{84}Po}, \ce{^{214}_{84}Po}, \ce{^{215}_{84}Po}, \ce{^{216}_{84}Po}, \ce{^{218}_{84}Po}, \ce{^{220}_{86}Rn}, \ce{^{222}_{86}Rn}, \ce{^{224}_{88}Ra}}, \ce{^{211}_{83}Bi}, \ce{^{219}_{86}Rn}, \ce{^{223}_{88}Ra}, \ce{^{255}_{100}Fm}\\
     & $0.12$ & \textcolor{red}{\ce{^{212}_{84}Po}, \ce{^{214}_{84}Po}, \ce{^{215}_{84}Po}, \ce{^{216}_{84}Po}, \ce{^{218}_{84}Po}, \ce{^{219}_{86}Rn}, \ce{^{220}_{86}Rn}, \ce{^{222}_{86}Rn}, \ce{^{224}_{88}Ra}}, \ce{^{211}_{83}Bi}\\
     & $0.14$ & \textcolor{red}{\ce{^{214}_{84}Po}, \ce{^{216}_{84}Po}, \ce{^{218}_{84}Po}}, \ce{^{211}_{83}Bi}, \ce{^{212}_{84}Po}, \ce{^{215}_{84}Po}, \ce{^{219}_{86}Rn}, \ce{^{220}_{86}Rn}, \ce{^{222}_{86}Rn}, \ce{^{224}_{88}Ra}, \ce{^{253}_{99}Es}, \ce{^{255}_{100}Fm}\\
     & $0.16$ & \textcolor{red}{\ce{^{212}_{84}Po}, \ce{^{214}_{84}Po}, \ce{^{216}_{84}Po}, \ce{^{218}_{84}Po}, \ce{^{220}_{86}Rn}, \ce{^{222}_{86}Rn}, \ce{^{224}_{88}Ra}}, \ce{^{211}_{83}Bi}, \ce{^{215}_{84}Po}, \ce{^{219}_{86}Rn}, \ce{^{255}_{100}Fm}\\
     & $0.18$ & \textcolor{red}{\ce{^{211}_{83}Bi}, \ce{^{212}_{84}Po}, \ce{^{214}_{84}Po}, \ce{^{215}_{84}Po}, \ce{^{216}_{84}Po}, \ce{^{218}_{84}Po}, \ce{^{219}_{86}Rn}, \ce{^{220}_{86}Rn}, \ce{^{222}_{86}Rn}, \ce{^{224}_{88}Ra}}\\
     & $0.21$ & \textcolor{red}{\ce{^{211}_{83}Bi}, \ce{^{212}_{84}Po}, \ce{^{214}_{84}Po}, \ce{^{215}_{84}Po}, \ce{^{216}_{84}Po}, \ce{^{218}_{84}Po}, \ce{^{219}_{86}Rn}, \ce{^{220}_{86}Rn}, \ce{^{222}_{86}Rn}, \ce{^{224}_{88}Ra}}\\ 
     & $0.24$ & \textcolor{red}{\ce{^{211}_{83}Bi}, \ce{^{212}_{84}Po}, \ce{^{214}_{84}Po}, \ce{^{215}_{84}Po}, \ce{^{216}_{84}Po}, \ce{^{218}_{84}Po}, \ce{^{219}_{86}Rn}, \ce{^{220}_{86}Rn}, \ce{^{222}_{86}Rn}, \ce{^{224}_{88}Ra}}\\
     & $0.28$ & \ce{^{211}_{83}Bi}, \ce{^{212}_{84}Po}, \ce{^{214}_{84}Po}, \ce{^{215}_{84}Po}, \ce{^{216}_{84}Po}, \ce{^{218}_{84}Po}, \ce{^{219}_{86}Rn}, \ce{^{220}_{86}Rn}, \ce{^{222}_{86}Rn}, \ce{^{224}_{88}Ra}\\
\enddata
\tablenotetext{}{Top contributing $\alpha$-decay nuclei that are common in at least 28 out of 32 nuclear physics input combinations are in red. Nuclei in black are common in at least 16 out of 32 nuclear physics input combinations.}
\end{deluxetable}

\newpage

\startlongtable
\begin{deluxetable}{ccc}
\tablecaption{\label{tabbmd} Key contributing $\beta$-decay reactions. }
\tablehead{\colhead{Time (d)} & \colhead{$Y_e$} & \colhead{$(Z,N)$}}
\startdata
\renewcommand{\arraystretch}{2.0}
 $1$ & $0.02$ & \textcolor{red}{\ce{^{128}_{51}Sb}, \ce{^{132}_{53}I}, \ce{^{133}_{53}I}, \ce{^{135}_{53}I}, \ce{^{135}_{54}Xe}, \ce{^{197}_{78}Pt}}, \ce{^{127}_{51}Sb}, \ce{^{129}_{51}Sb}, \ce{^{129}_{52}Te}, \ce{^{132}_{52}Te}, \ce{^{131}_{53}I}, \ce{^{143}_{58}Ce}, \ce{^{193}_{76}Os}\\
     & $0.12$ & \textcolor{red}{\ce{^{132}_{53}I}, \ce{^{133}_{53}I}, \ce{^{135}_{53}I}, \ce{^{135}_{54}Xe}, \ce{^{197}_{78}Pt}}, \ce{^{127}_{51}Sb}, \ce{^{128}_{51}Sb}, \ce{^{129}_{51}Sb}, \ce{^{129}_{52}Te}, \ce{^{132}_{52}Te}, \ce{^{131}_{53}I}, \ce{^{141}_{57}La}, \ce{^{193}_{76}Os}\\
     & $0.14$ & \textcolor{red}{\ce{^{132}_{53}I}, \ce{^{133}_{53}I}, \ce{^{135}_{53}I}, \ce{^{135}_{54}Xe}}, \ce{^{127}_{51}Sb}, \ce{^{128}_{51}Sb}, \ce{^{129}_{51}Sb}, \ce{^{129}_{52}Te}, \ce{^{132}_{52}Te}, \ce{^{131}_{53}I}\\
     & $0.16$ & \textcolor{red}{\ce{^{128}_{51}Sb}, \ce{^{132}_{53}I}, \ce{^{133}_{53}I}, \ce{^{135}_{54}Xe}, \ce{^{197}_{78}Pt}}, \ce{^{127}_{51}Sb}, \ce{^{129}_{51}Sb}, \ce{^{129}_{52}Te}, \ce{^{131}_{53}I} \ce{^{135}_{53}I}, \ce{^{240}_{93}Np}\\
     & $0.18$ & \textcolor{red}{\ce{^{128}_{51}Sb}, \ce{^{132}_{53}I}, \ce{^{133}_{53}I}, \ce{^{193}_{76}Os}, \ce{^{197}_{78}Pt}, \ce{^{208}_{81}Tl}, \ce{^{240}_{93}Np}}\\
     & & \ce{^{129}_{51}Sb}, \ce{^{129}_{52}Te}, \ce{^{135}_{53}I}, \ce{^{135}_{54}Xe}, \ce{^{200}_{79}Au}, \ce{^{212}_{82}Pb}\\
     & $0.21$ & \textcolor{red}{\ce{^{132}_{53}I}, \ce{^{133}_{53}I}, \ce{^{135}_{53}I}, \ce{^{135}_{54}Xe}, \ce{^{157}_{63}Eu}, \ce{^{184}_{73}Ta}, \ce{^{189}_{75}Re}, \ce{^{193}_{76}Os}, \ce{^{197}_{78}Pt}}\\
     & & \ce{^{128}_{51}Sb}, \ce{^{131}_{53}I}, \ce{^{140}_{57}La}, \ce{^{151}_{61}Pm}, \ce{^{159}_{64}Gd}, \ce{^{166}_{67}Ho}, \ce{^{171}_{68}Er}, \ce{^{172}_{68}Er}, \ce{^{187}_{74}W}\\
     & $0.24$ & \textcolor{red}{\ce{^{128}_{51}Sb}, \ce{^{129}_{51}Sb}, \ce{^{129}_{52}Te}, \ce{^{132}_{52}Te}, \ce{^{132}_{53}I}, \ce{^{133}_{53}I}, \ce{^{135}_{53}I}, \ce{^{135}_{54}Xe}, \ce{^{193}_{76}Os}}\\
     & & \ce{^{131}_{53}I}, \ce{^{140}_{57}La}, \ce{^{151}_{61}Pm}, \ce{^{156}_{62}Sm}, \ce{^{157}_{63}Eu}, \ce{^{159}_{64}Gd}, \ce{^{166}_{67}Ho}, \ce{^{184}_{73}Ta}\\
     & $0.28$ & \textcolor{red}{\ce{^{132}_{53}I}, \ce{^{133}_{53}I}}, \ce{^{128}_{51}Sb}, \ce{^{132}_{52}Te}, \ce{^{131}_{53}I}, \ce{^{135}_{53}I}, \ce{^{135}_{54}Xe}\\
 $8$ & $0.02$ & \textcolor{red}{\ce{^{132}_{52}Te}, \ce{^{132}_{53}I}, \ce{^{133}_{54}Xe}, \ce{^{140}_{57}La}}, \ce{^{125}_{50}Sn}, \ce{^{127}_{51}Sb}, \ce{^{127}_{52}Te}, \ce{^{131}_{53}I}, \ce{^{140}_{56}Ba}, \ce{^{156}_{63}Eu}\\
     & $0.12$ & \textcolor{red}{\ce{^{132}_{52}Te}, \ce{^{132}_{53}I}, \ce{^{133}_{54}Xe}, \ce{^{140}_{57}La}}, \ce{^{125}_{50}Sn}, \ce{^{127}_{51}Sb}, \ce{^{127}_{52}Te}, \ce{^{131}_{53}I}, \ce{^{140}_{56}Ba}, \ce{^{156}_{63}Eu}, \ce{^{166}_{67}Ho}\\
     & $0.14$ & \textcolor{red}{\ce{^{132}_{53}I}}, \ce{^{125}_{50}Sn}, \ce{^{127}_{51}Sb}, \ce{^{132}_{52}Te}, \ce{^{131}_{53}I}, \ce{^{133}_{54}Xe}, \ce{^{140}_{56}Ba}, \ce{^{140}_{57}La}\\
     & $0.16$ & \textcolor{red}{\ce{^{132}_{52}Te}, \ce{^{131}_{53}I}, \ce{^{132}_{53}I}}, \ce{^{125}_{50}Sn}, \ce{^{127}_{51}Sb}, \ce{^{127}_{52}Te}, \ce{^{133}_{54}Xe}, \ce{^{140}_{56}Ba}, \ce{^{140}_{57}La}, \ce{^{234}_{91}Pa}, \ce{^{246}_{95}Am}\\ 
     & $0.18$ & \textcolor{red}{\ce{^{127}_{51}Sb}, \ce{^{132}_{52}Te}, \ce{^{131}_{53}I}, \ce{^{132}_{53}I}, \ce{^{208}_{81}Tl}, \ce{^{234}_{91}Pa}, \ce{^{246}_{95}Am}}, \ce{^{125}_{50}Sn}, \ce{^{127}_{52}Te}, \ce{^{133}_{54}Xe}\\
     &  &\ce{^{140}_{56}Ba}, \ce{^{140}_{57}La}, \ce{^{193}_{76}Os}, \ce{^{199}_{79}Au}, \ce{^{202}_{79}Au}, \ce{^{207}_{81}Tl}, \ce{^{211}_{82}Pb}, \ce{^{212}_{82}Pb}, \ce{^{214}_{82}Pb}, \ce{^{247}_{94}Pu}\\ 
     & $0.21$ & \textcolor{red}{\ce{^{132}_{53}I}, \ce{^{140}_{57}La}, \ce{^{156}_{63}Eu}, \ce{^{166}_{67}Ho},\ce{^{172}_{69}Tm}, \ce{^{191}_{76}Os}, \ce{^{193}_{76}Os}} \\
     &  & \ce{^{132}_{52}Te}, \ce{^{131}_{53}I}, \ce{^{140}_{56}Ba}, \ce{^{166}_{66}Dy}, \ce{^{188}_{75}Re}\\ 
     & $0.24$ & \textcolor{red}{\ce{^{131}_{53}I}, \ce{^{132}_{53}I}, \ce{^{140}_{57}La}, \ce{^{166}_{67}Ho}}, \ce{^{132}_{52}Te}, \ce{^{133}_{54}Xe}, \ce{^{140}_{56}Ba}, \ce{^{156}_{63}Eu},\ce{^{172}_{69}Tm}\\
     & $0.28$ & \textcolor{red}{\ce{^{131}_{53}I}, \ce{^{132}_{53}I}}, \ce{^{132}_{52}Te}
\enddata
\tablenotetext{}{Top contributing $\beta$-decay nuclei that are common in at least 28 out of 32 nuclear physics input combinations are in red. Nuclei in black are common in at least 16 out of 32 nuclear physics input combinations.}
\end{deluxetable}
\end{document}